\documentclass{aa}
\usepackage{booktabs}
\usepackage{graphicx}
\usepackage[varg]{txfonts}
\usepackage[colorlinks=true, citecolor=blue]{hyperref}
\usepackage{multirow}
\usepackage{multicol}
\usepackage{natbib}
\usepackage[normalem]{ulem}
\usepackage{lastpage}

\bibpunct{(}{)}{;}{a}{}{,}

\begin{document} 

   \title{The ALPINE-ALMA [CII] Survey: Modelling ALMA and JWST lines to constrain the interstellar medium of $z\sim 5$ galaxies}
   
   \subtitle{Connecting UV, Optical, and FIR line emission}

   \author{E. Veraldi\inst{1,2}\fnmsep\thanks{enrico.veraldi2@unibo.it}, L. Vallini\inst{3}, F. Pozzi\inst{2,3}, F. Esposito\inst{2,3}, M. Bethermin\inst{4,5}, M. Boquien\inst{6}, A. Faisst\inst{7},\\M. Ginolfi\inst{8}, R. Gobat\inst{9}, C. Gruppioni\inst{3}, N. Hathi\inst{10}, E. Ibar\inst{9}, J. Molina\inst{9}, F. Rizzo\inst{11,12,15},\\M. Romano\inst{16,13,14}, G. Zamorani\inst{3}
          }
          
   \institute{
            Scuola Internazionale Superiore Studi Avanzati (SISSA), Physics Area, Via Bonomea 265, 34136 Trieste, Italy
            \and
            Dipartimento di Fisica e Astronomia, Universitá degli Studi di Bologna, Via P. Gobetti 93/2, I-40129 Bologna, Italy
            \and
            Osservatorio di Astrofisica e Scienza dello Spazio (INAF–OAS), Via P. Gobetti 93/3, I-40129 Bologna, Italy
            \and
            Université de Strasbourg, CNRS, Observatoire astronomique de Strasbourg, UMR 7550, 67000 Strasbourg, France
            \and
            Aix Marseille Univ, CNRS, CNES, LAM, Marseille, France
            \and
            Université Côte d’Azur, Observatoire de la Côte d’Azur, CNRS, Laboratoire Lagrange, 06000, Nice, France
            \and
            Caltech/IPAC, 1200 E. California Blvd. Pasadena, CA 91125, USA
            \and
            Dipartimento di Fisica e Astronomia, Università di Firenze, Via G. Sansone 1, 50019, Sesto Fiorentino (Firenze), Italy
            \and
            Instituto de Física y Astronomía, Universidad de Valparaíso, Avda. Gran Bretaña 1111, Valparaíso, Chile
            \and
            Space Telescope Science Institute, 3700 San Martin Drive, Baltimore, MD 21218, USA
            \and
            Cosmic Dawn Center (DAWN), Copenhagen, Denmark
            \and
            Niels Bohr Institute, University of Copenhagen, Jagtvej 128, 2200 Copenhagen, Denmark
            \and
            National Centre for Nuclear Research, ul. Pasteura 7, 02-093 Warsaw, Poland
            \and
            INAF - Osservatorio Astronomico di Padova, Vicolo dell'Osservatorio 5, I-35122, Padova, Italy
            \and
            Kapteyn Astronomical Institute, University of Groningen, Landleven 12, 9747 AD, Groningen, The Netherlands
            \and Max-Planck-Institut für Radioastronomie, Auf dem Hügel 69, 53121 Bonn, Germany 
            }
             
   \date{Received 14 June 2024 / Accepted 7 November 2024}

  \abstract
    {}
    {In this work we devise a model for estimating ultraviolet (UV) and optical line emission (i.e. CIII] $1909\rm \text{\AA}$, H$\beta$, [OIII] $5007\rm \text{\AA}$, H$\alpha$, [NII] $6583\rm \text{\AA}$) tracing HII regions in the interstellar medium (ISM) of a subset of galaxies at $z\sim4-6$ from the ALMA Large Programme ALPINE. The aim is to investigate the combined impact of binary stars in the stellar population along with the presence of an abrupt quenching in the Star Formation History (SFH) on the line emission.
    This is crucial for understanding the ISM's physical properties in the universe's earliest galaxies and identifying new star formation tracers in high-$z$ galaxies.}
    {The model simulates HII+Photodissociation Region (PDR) complexes by performing radiative transfer through 1D slabs characterised by gas density ($n$), ionisation parameter ($U$), and metallicity ($Z$). The model takes into account of: \textit{(a)} the heating from star formation (SF), whose spectrum has been simulated with the Starburst99 and Binary Population and Spectral Synthesis (BPASS) to quantify the impact of binary stars; \textit{(b)} a constant, exponentially declining, and quenched SFH.
    For each galaxy we select from our CLOUDY models the theoretical ratios between the [CII] line emission, tracing PDRs, and nebular lines from HII regions. These ratios have been then used to derive the expected optical/UV lines from the observed [CII]. }
    {We find that binary stars have a strong impact on the line emission after quenching, by keeping the UV photon flux higher for longer time. This is relevant in maintaining the free electron temperature and ionised column density in HII regions unaltered up to 5 Myr after quenching. Furthermore, we constrain the ISM  properties of our subsample, finding a low ionisation parameter $\log U{\approx}-3.8\pm 0.2$, and moderate/high densities of $\log(n/\rm cm^{-3}){\approx}2.9\pm 0.6$. Finally, we derive UV/optical line luminosities-SFR relations ($\log(L_{line}/\rm erg\, s^{-1})=\alpha\log(\rm SFR/M_{\odot}\, yr^{-1})+\beta$) for different burstiness parameter ($k_s$) values. We find that in the fiducial BPASS model, the relations have negligible SFH dependence but significantly depend on the $k_s$ value, while in the SB99 case, the dominant dependence is on the SFH. We propose their potential use for characterising the burstiness of galaxies at high-$z$.}
    {}

   \keywords{HII regions -- photon-dominated region (PDR) -- Galaxies: ISM -- Galaxies: star formation -- Galaxies: high-redshift}

\titlerunning{ISM properties in ALPINE galaxies}
\authorrunning{E.Veraldi et al.$\,$}
\maketitle

\section{Introduction}

The Epoch of Reionisation (EoR; redshift $6<z<30$) is a fundamental era of the cosmic history during which the ionising photons produced by star formation (SF) in the first galaxies affected the ionisation state of the Universe \citep[][and reference therein]{Dayal+18, Robertson2022}. For this reason, shedding light on the SF process during the transition period immediately after the EoR and before the beginning of cosmic noon, when the bulk of the stellar mass of the Universe was formed, is of utmost interest.
Star formation is shaped by the properties of the gas collapsing into stars \citep{Tacconi+20}, therefore constraining the properties of the interstellar medium (ISM) of galaxies at $z\sim 4-6$ is crucial. 

On the theoretical side, ISM models and simulations seem to agree that the prevailing physical conditions within galaxies at the end of the EoR were different from those in the local Universe \citep{Vallini+15,Katz+17, Katz+22,Lagache+18,Pallottini+19, Pallottini+22,Arata+20,Lupi+20}. In particular, zoom-in simulations and analytical models developed to interpret neutral gas tracers \citep[e.g. ][]{Vallini+15,Lagache+18,Pallottini+22}, ionised gas tracers \citep[e.g. ][]{Moriwaki+18,Arata+20,Katz+22}, and dust continuum emission \citep[e.g. ][]{Behrens+18,DiCesare+23} find high turbulence \citep[e.g. ][]{Kohandel+20}, strong radiation fields \citep[e.g. ][]{Katz+22}, high densities \citep[e.g. ][]{Nakazato+23}, and warm dust temperatures \citep[e.g. ][]{Sommovigo+21, Vallini+21}, to be common on sub-kpc scales in the ISM of star-forming high-$z$ galaxies. These findings have to be tested against high-resolution observations.

In the last ten years, the Atacama Large Millimetre/submillimeter Array (ALMA) revolutionised the study of neutral and molecular gas phases in high-$z$ sources, by opening a window on the (rest frame) far-infrared (FIR) lines \citep{Carilli2013}.
The $^2P_{3/2}\rightarrow ^2P_{1/2}$ fine structure transition of the ionised carbon (\rm{[CII]}) at $158 \, \rm{\mu m}$, being the most luminous line in the FIR band \citep[][]{stacey2010}, has become the most widely targeted transition by ALMA in the high-$z$ Universe \citep[e.g.][]{Maiolino,Carniani,Matthee+19}. The $C^0$ ionisation potential ($11.26 \rm\, eV$) is lower than that of neutral hydrogen ($13.6 \, \rm{eV}$) making C$^+$ abundant in the diffuse neutral medium and in dense PhotoDissociation Regions \citep[PDRs;][for a recent review]{Wolfire22} associated with molecular clouds. PDRs are defined as regions where the far-ultraviolet (FUV, $6-13.6 \, \rm{eV}$) radiation from OB stars, energetic enough to dissociate molecules but not to ionise hydrogen, drives the chemistry and thermal balance of the gas.

In addition to [CII] campaigns targeting single galaxies, two ALMA surveys have been carried out on large samples: the ALMA Large Programme to Investigate \rm{[CII]} at Early Times \citep[ALPINE,][at $z=4-5$]{Fevre, Faisst, Bethermin+20}, and the Reionisation Era Bright Emission Line Survey \citep[REBELS,][at $z=6-7$]{Bouwens}. Due to the balance between [CII] cooling and heating from star formation, \citet{DeLooze} first established the [CII]-SFR relation on both galactic scales and on a spatially resolved level at $z=0$. On galactic scales, this relationship seems to hold both in ALPINE \citep[e.g.][]{Schaerer, Romano+22} and REBELS \citep{Bouwens} and up to $z=7$ \citep{Carniani+18,Matthee+19}, albeit with larger scatter. In addition to the SFR, the luminosity of [CII] also correlates with the gas mass \citep{Zanella} due to the widespread C$^+$ abundance in the neutral and molecular phases.\\

The ionised gas in galaxies is instead usually traced with optical and ultraviolet (UV) nebular lines \citep[e.g. \rm{[OII], H$\beta$, [OIII], [NII], H$\alpha$, [SII]};][and references therein]{kewley2019}. By leveraging line ratios diagnostics (e.g. $R2$, $O3O2$, $R3$, $R23$) and well-established correlations between their luminosity and the SFR (e.g. $\rm L_{H\alpha}$-SFR and $\rm L_{[OII]}$-SFR, \citealt{Kennicutt}) it is possible to characterise the ionised gas temperature \citep[][]{Schaerer+22}, metallicity \citep[e.g.][]{curti2023,venturi2024}, and density \citep{isobe2023}. Observations of the ionised gas in high-$z$ galaxies have historically been performed with ground-based facilities (e.g. VLT and Keck telescopes) by targeting rest-frame optical/UV emission lines up to $z\sim 3$, and with the Hubble Space Telescope. Large programmes like the VIMOS Survey in the CANDELS UDS and CDFS fields \citep[VANDELS, ][]{McLure+18}; the Keck Baryonic Structure Survey \citep[KBSS, ][]{Steidel+14} or the MOSFIRE Deep Evolution Field (MOSDEF) survey \citep{Kriek+15} allowed a first analysis of the evolution of ionised gas properties.
These campaigns enabled the analysis of the validity of commonly used line diagnostics (e.g. the BPT diagram, \citealt{Baldwin+91}), and the derivation of the underlying ionised gas properties in galaxies beyond the local Universe. Overall, such works suggest a complex interplay between star
formation, gas kinematics, and chemical enrichment in relatively
young galaxies at $z\sim3$ \citep{Llerena23}.
However, it is only in the last two years that, thanks to the unprecedented sensitivity and resolution of the James Webb Space Telescope (JWST), the ionised gas has become accessible well within the EoR with deep spectro/photometric campaigns such as the JWST Advanced Deep Extragalactic Survey \citep[JADES;][]{Eisenstein,Bunker+23, Rieke+23, Tacchella+23}, the Cosmic Evolution Early universe Release Science Survey \citep[CEERS;][]{Yang+23,Backhaus+24}, or the Public Release IMaging for Extragalactic Research \citep[PRIMER;][]{Dunlop+21,Donnan+24}. JWST is making it possible to test commonly used relations that have been calibrated in local galaxies, and to link the HII region properties with those of the neutral gas traced with ALMA in [CII]. 

The goal of this work is to link the luminosity of nebular lines tracing ionised gas, to that of [CII] tracing neutral gas, and clarifying the effect of binary stars and quenching episodes in star formation on the line luminosity and ratios.
In particular, we focused on a subsample of ALPINE galaxies at $z\approx 4-6$.
A JWST proposal has been accepted (PI: A. {Faisst}, Cycle 2) to observe 18 representative ALPINE galaxies between $4.4<z<5.7$ with NIRSpec, and obtain all major optical lines (\rm{[OII], H$\beta$, [OIII]$5007\rm \text{\AA}$, [NII], H$\alpha$ and [SII]}) for which we will provide luminosity predictions based on the known properties of ALPINE galaxies inferred from the [CII] and dust continuum observations.

Works based on pre-JWST data suggest that ALPINE$-$like galaxies \citep[][$z\sim 5$]{Faisst+16}, and a subsample of 10 ALPINE sources \citep{Vanderhoof+22} to be characterised by $Z=0.5 \,\rm Z_{\odot}$, while ALMA [CII] and continuum detections allowed the characterisation of the morphology and distribution of the cold gas and dust \citep{Fujimoto, Pozzi}, the measure of the gas mass \citep[][]{DZ}, and of the \rm{[CII]}-SFR relation \citep[e.g. ][]{Bethermin,Schaerer,DZ,Romano+22}) in ALPINE. The upcoming JWST data, complemented with the models presented in this paper, will provide us with the unique opportunity of having a UV-to-submillimeter benchmark sample with well-characterised multiphase properties.\\

This paper is organised as follows: in Section \ref{sec:modelling} we present the model. In Section \ref{sec:Model_appl} we describe the method used to apply the model to the selected ALPINE subsample. 
In Section \ref{sec:Results} we discuss the results, in particular predictions about UV/optical lines (\rm{[OIII]$5007\rm \text{\AA}$}, \rm{[NII]$6583\rm \text{\AA}$}, $\rm{H\alpha}$, $H\beta$ and \rm{CIII]$1909\rm \text{\AA}$})-SFR relations, and the impact of binary stars and quenching on these predictions.
In Section \ref{sec:conclusion} we present our conclusions.

\section{Model outline}\label{sec:modelling}

The line emission from the ionised and neutral gas in the [CII]-detected galaxies considered in this work is computed by combining the radiative transfer (RT) calculations performed with \textsc{CLOUDY} \citep[version C22.02,][]{Cloudy+23} with the analytical treatment of the [CII] emission presented in \citet{Ferrara}. More precisely, \textsc{CLOUDY} is exploited to derive the line emissivity as a function of the radiation field, gas density, and metallicity (see Sec. \ref{sec:CLOUDY} for details), while the \citet{Ferrara} model allows us to derive the gas density and ionisation parameter from the measured SFR surface density ($\Sigma_{\rm SFR}$), thus enabling the selection of tailored \textsc{CLOUDY} models for each galaxy from our simulation library (see Sec. \ref{sec:deriv_pred} for details).

\subsection{CLOUDY modelling}\label{sec:CLOUDY}

\textsc{CLOUDY} \citep{Ferland} is a non-local thermodynamic
equilibrium photoionisation code designed to simulate astrophysical environments and their emerging spectra.
We employ \textsc{CLOUDY v22.02} \citep{Cloudy+23} to model the ISM conditions within our galaxy sample. 
We focus on the HII regions, created by the extreme ultraviolet (EUV, $h\nu>13.6\, \rm{eV}$) photons, and on the PDRs produced by far-ultraviolet (FUV, $6<E<13.6\, \rm{eV}$) radiation from newly formed OB stars. 
\textsc{CLOUDY} simulates HII region + PDR complexes by performing the RT through a 1-D gas slab for which we assume constant hydrogen number density ($n$), ionisation parameter ($U$)\footnote{The ionisation parameter $U$ is defined as the dimensionless ratio of hydrogen-ionising photon flux ($\phi$) to total-hydrogen density ($n$) at the illuminated face of the gas slab:
\begin{equation}\label{eq:ionisation_parameter}
    U=\frac{\phi}{nc}.
\end{equation}}, and gas metallicity ($Z$).
We produce a grid of \textsc{CLOUDY} models by varying the hydrogen number density in the range $\log(n/\rm cm^{-3})=[0.5,7]$ ($0.5$ dex steps), the ionisation parameter in the range $\log U=[-4.5,-0.5]$ ($0.5$ dex steps), and the gas metallicity in the range $Z/Z_{\odot}=[0.15,0.55]$ assuming $0.1$ dex steps (see Table \ref{tab:CLOUDY_grid_values}).

\begin{table}[htbp]
    \centering
    \caption{Parameter grid characterising the developed \textsc{CLOUDY} models.}
    \label{tab:CLOUDY_grid_values}
    \begin{tabular}{ll}
    \toprule
    \toprule
        $\log(n/\rm cm^{-3})$ & $[0.5,7]$, $0.5$ dex steps\\
        $\log(U)$ & $[-4.5,-0.5]$, $0.5$ dex steps\\
        $Z\, \mathrm{(Z_{\odot})}$ & $[0.15,0.55]$, $0.1$ dex steps\\
    \bottomrule  
    \end{tabular}
\end{table}

The abundances of metals at solar metallicity are set to match those by \citet{Grevesse}. We assume that all of them linearly scale with $Z$, thus their relative ratios is constant with $Z$. We include the \textsc{CLOUDY} default ISM grain distribution \citep{Mathis+77}, and we assume the dust-to-gas ratio to linearly scale with $Z$ \citep{Baldwin+91,WeingartnerDraine+01,vanHoof+04,Weingartner+06}.
We account for the Cosmic Microwave Background (CMB) \citep{Mather+99} at the redshift of the ALPINE galaxies ($z\approx4-5.5$, $T_{\rm CMB} = 2.7(1+z)\, \rm K$). We set the microturbulence gas velocity to $v_{turb}=1.5 \, \rm km \, s^{-1}$, and we include the effect of cosmic rays (CRs) by assuming an ionisation rate $\xi_{CR}=2\times10^{-16}\, \rm s^{-1} $ \citep{Indriolo}. Turbulence affects the shielding and pumping of the lines, while CRs must be included because their impact on the gas heating and chemistry is not negligible when the calculation extends into the molecular region.
We stop the \textsc{CLOUDY} simulations when we reach a total hydrogen column density $N_{\rm H}=10^{22}\,\rm cm^{-2}$, in order to fully sample both the HII region and the neutral gas in the PDR out to the edge of the fully molecular part. In all our simulations, we assume that a stellar Spectral Energy Distribution (SED), normalised with $U$, illuminates the gas slab from one side. In what follows, we provide an extensive discussion regarding the SED modelling.

\subsection{Incident radiation field}\label{sec:SB99/BPASS}
Given that the major fraction of the ALPINE galaxies do not show any significant AGN activity (see Section \ref{sec:dataset} for details), our model assume that the only source of radiation striking the gas slab is that produced by star formation. From \citet{Faisst} we consider the average stellar age within the ALPINE sample, $t_{*}=224\, \rm Myr$,
estimated from the SED fitting, assuming a Chabrier Initial Mass Function (IMF).
We test three possible star formation histories (SFHs):
\textit{I)} continuous star formation through $\Delta t =224 \rm \,Myr$, \textit{II)} exponentially declining SFH ($\propto e^{-\tau}$) as assumed by \citet{Faisst} to fit the majority of the ALPINE SEDs, \textit{III)} constant SFH with an instantaneous quenching episode occurred at $\Delta t_{q} = 5\, \rm Myr$ prior the observation.

The latter SFH has been considered to assess the impact of binary stars in sustaining emission shortly after SF cessation.
Note that in all SFHs we do not account for stars older than $224\, \rm{Myr}$. 
Nevertheless, even if old stellar populations account for a major fraction of the stellar mass, they do not significantly affect the ionising photon budget; thus their emission is not relevant for the CLOUDY modelling.

\begin{figure}[htbp]
    \centering
    \resizebox{\hsize}{!}{\includegraphics{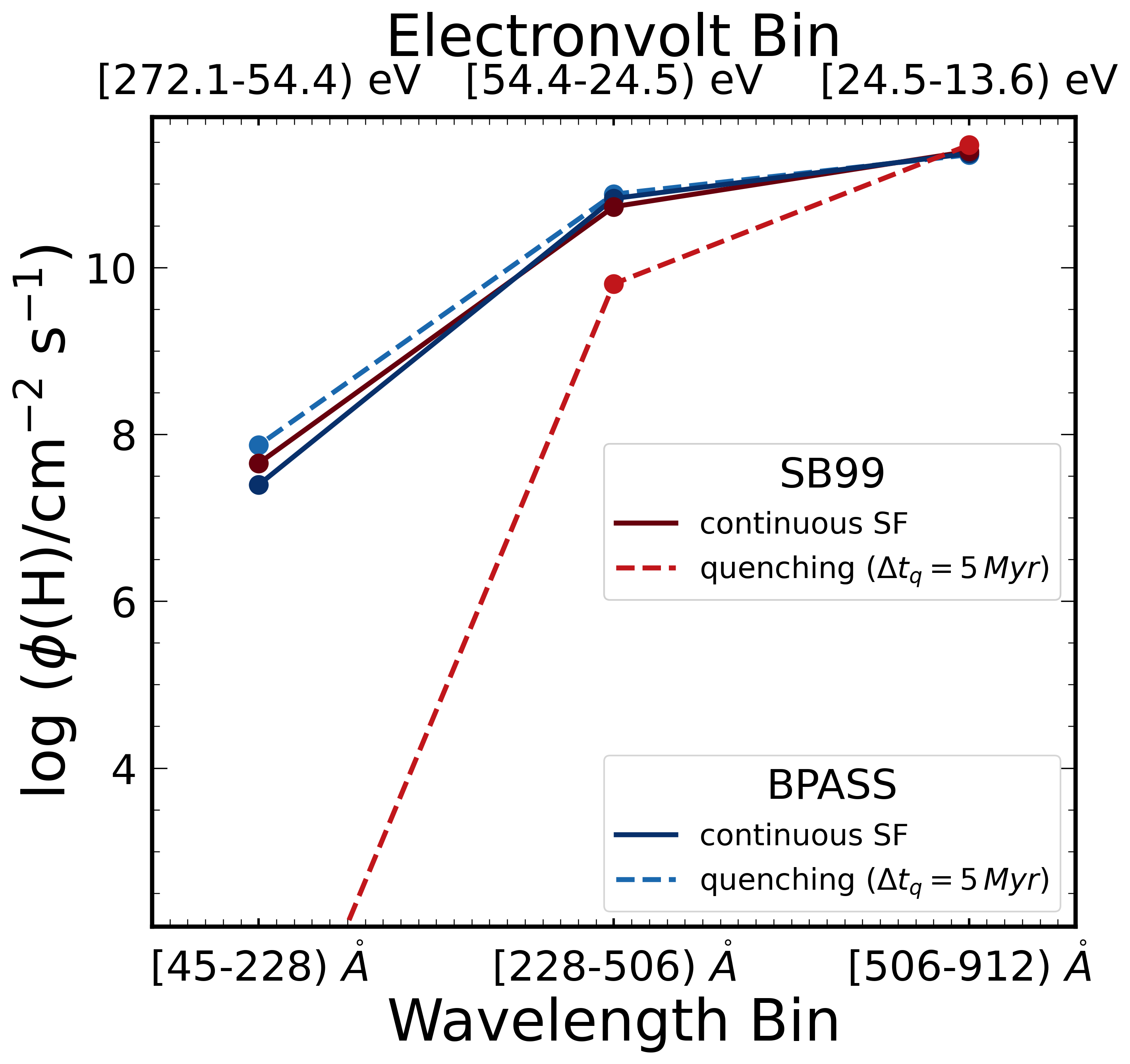}}
    \caption{The incident ionising flux $\phi(H)$ per energy/wavelength bin (\textsc{CLOUDY} model $\log U=-1.00$, $\log n=2$, $Z=0.55\, Z_{\odot}$) for SB99 (red lines) and BPASS (blue lines). The different SFH models are represented with solid (no quenching) and dashed ($\Delta{t}_q=5$ Myr) lines, respectively.}
    \label{fig:phi_main}
\end{figure}
For each SFH we build the composite SED by summing single stellar populations (SSPs) computed with Starburst99 v7.0.1 \citep[SB99 hereafter]{SB99}, and BPASS v2.2.1 \citep[BPASS hereafter]{stanway}. SB99 is a spectral synthesis code that returns the SED of SSPs without binary stars. For building the SSPs, we choose Geneva evolutionary tracks with standard mass loss \citep[][]{Schaerer+93}, to generate a SED with a stellar metallicity $Z_{*}=0.4\, \rm{Z_{\odot}}$ \citep[][]{Vanderhoof+22}, and a Chabrier IMF \citep[][]{Chabrier+03} (see Table \ref{tab:BPASS_SB99_setting}).
The stellar metallicity is consistent with that inferred for the ALPINE sample and similar galaxies \citep{Faisst+16,Vanderhoof+22}. Fixing $Z_*$ can affect the hardness of the spectra, leading to uncertainties in the estimated value of the ionisation parameter $U$. The upcoming JWST data will alleviate this problem by constraining $Z_*$ to a better precision.

BPASS is a suite of synthetic stellar population and binary stellar evolution models. Unlike SB99, BPASS includes the effect of mass transfer in binary systems. That provides additional fuel to burn as an accretion disc \citep[][]{Kartunnen}, and this leads to a greater emission in FUV and EUV for a longer time than if there were no binaries \citep{Eldridge2022}. Many authors \cite[e.g.][]{Stanaway+14,Jaskot16,Xiao+18,Xiao+19} pointed out that this might have a non-negligible effect on the resulting HII region properties and on the related line emission. For this reason, we test the effect of BPASS SEDs on our \textsc{CLOUDY} models by comparing them with those obtained assuming the same SED but computed with SB99. In line with the SB99 modelling, we set $Z_{*}=0.4\, \mathrm{Z_{\odot}}$ and IMF parameters listed in Table \ref{tab:BPASS_SB99_setting}. BPASS SSPs use BPASS-developed evolutionary tracks.  \\

For a given SFH, the radiation emitted by the composite stellar population, $F(\lambda)$ is computed using the following equation:
\begin{equation}\label{eq:composite_stellar_pop}
    F(\lambda)=\psi_0 f_0(\lambda) \Delta t_0 + \sum_{i=1}^{224\, \rm Myr} \psi_i f_i(\lambda)\Delta t_i
\end{equation}

where $i$ denotes different SSPs of different ages, $f_i$ is the flux of the SSPs in the corresponding age bin ($\Delta t_i$), and $\psi_i$ is the instantaneous SFR in the $i$-bin.
For the continuous star formation mode, we assume $\log(\rm SFR/M_{\odot}\, yr^{-1}) = 1.7$, namely, the average value in the ALPINE sample \citep{Faisst}. For the SFH with quenching, we stop the sum of the SSPs (see Eq. \ref{eq:composite_stellar_pop}) at the age assumed for quenching. For the exponentially decreasing SFH, we assume $\psi(t)\propto e^{-\tau}$ with $\tau=0.3\, \rm Gyr$, that is the best-fit value for the majority of the ALPINE galaxies.
We normalise $\psi$ so that $\log(\rm SFR/M_{\odot}\, yr^{-1}) \approx 1.7$ at $224\, \rm{Myr}$.

\begin{table}[htbp]
    \centering
    \caption{SSP settings for BPASS and SB99 SED modelling.}
    \label{tab:BPASS_SB99_setting}
    \begin{tabular}{ccccc}
    \toprule
    \toprule
           {$M_{*}$} & {IMF exponent (a)} & {IMF exponent (b)} & {$M_{\star}$ range}  \\
          {$\mathrm{[M_{\odot}]}$} & {[0.1-0.5 $\mathrm{M_{\odot}}$]} & {[0.5-300 $\mathrm{M_{\odot}}$]} & {$\mathrm{[M_{\odot}]}$}  \\
    \midrule
      $10^6$ & $-1.30$ & $-2.35$ & $0.1-300$\\
    \bottomrule
    \end{tabular}
    \tablefoot{The two IMF exponents are valid in the reported range of masses. An IMF constructed with these exponents exhibits behaviour similar to a Chabrier one \citep{KE12}.
    The first column, $M_{*}$, is the mass of stars formed in each starburst.
    }
\end{table}

\subsection{Impact of SB99 vs BPASS on the HII region properties}\label{sec:SB99vsBPASS}

As outlined in Section \ref{sec:SB99/BPASS}, in our work we computed the SED for three SFHs.
The exponentially declining star formation history and the continuous one return SEDs with comparable fluxes in all the energy bins (see Appendix \ref{sec:exp_decl}). For this reason, in the remainder of this work we will only consider the continuous SFH and that characterised by the quenching episode 5 Myr prior observation. 

We start by analysing differences in the ionising-photon budget at fixed SFH between SB99 and BPASS spectra. In Figure \ref{fig:phi_main} we show the incident H-ionising photon flux at the gas slab, $\phi$, in bins with decreasing energy (A $[272.1-54.4)\rm \, eV$, B $[54.4-24.5)\rm \, eV$, C $[24.5-13.6)\rm \, eV$) for a fixed ionisation parameter $\log U=-1$ as computed by CLOUDY.

The impact on the ionising photon budget from binary stars is negligible for continuous SF (solid lines), hence SB99 and BPASS models are almost identical. In fact, stars are continuously formed and the youngest among them dominate the ionising photon flux in all the energy bins. The UV photons from binary systems are therefore negligible.
\begin{figure*}[htbp]
    \centering
    \includegraphics[width=16cm]{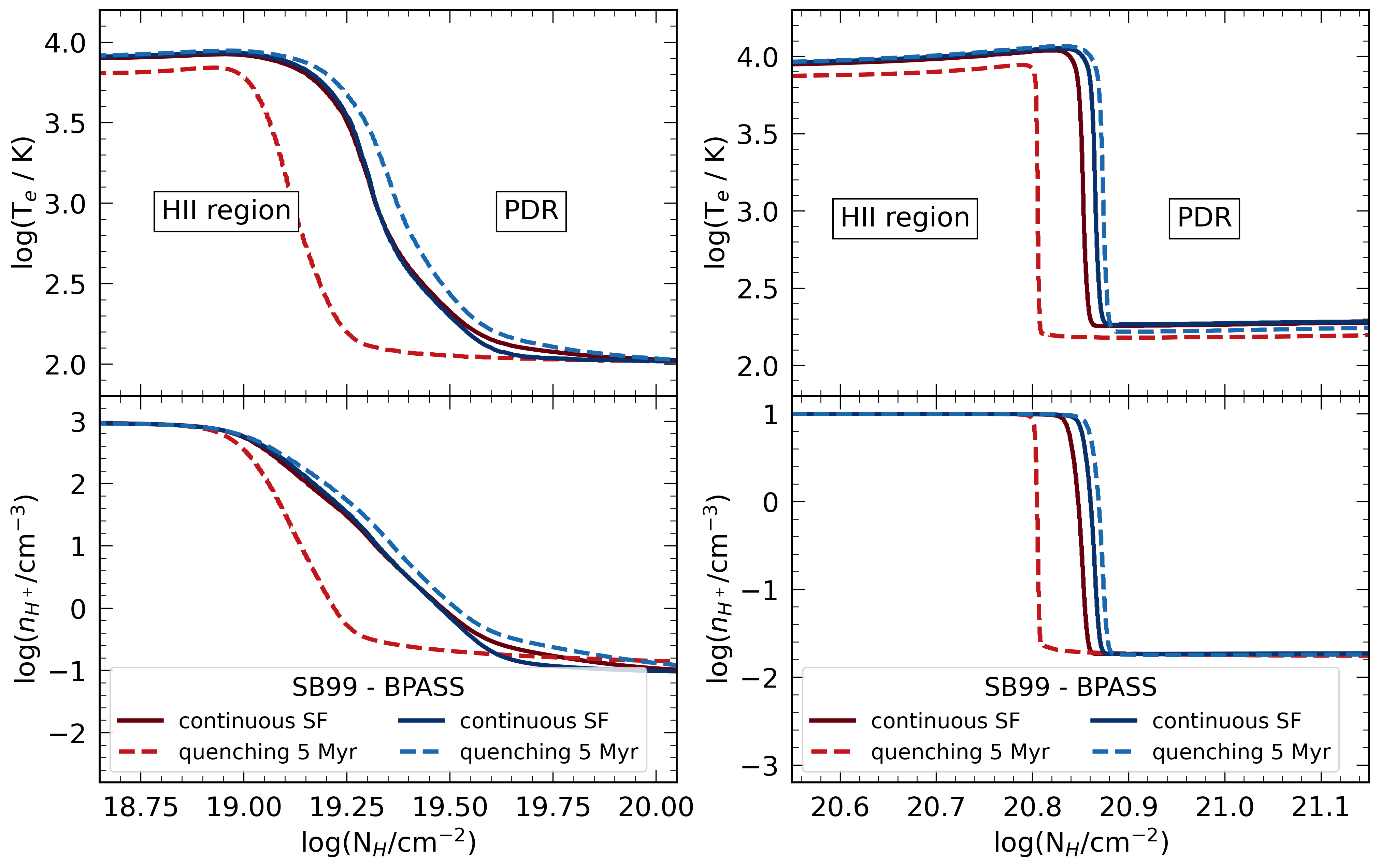}
    \caption{The free electron temperature ($T_e$, upper panels) and H$^+$ number density ($n_{H^+}$, lower panels) as a function of the hydrogen column density ($N_H$). \textsc{CLOUDY} models $\log U=-4.00$, $\log n=3.00$, $Z=0.55\, Z_{\odot}$ (left), $\log U=-2.00$, $\log n=1.00$, $Z=0.55\, Z_{\odot}$  (right). 
    These models represent typical ISM conditions for the ALPINE galaxies (Table \ref{table:ALPINE_data}).
    Solid and dashed lines represent the temperature and density profiles obtained for different SFHs. Red colour palette lines represent SB99 models, while blue colour palette lines represent BPASS models.}
\label{fig:temp}
\end{figure*}

We observe instead differences if the SFH is characterised by a quenching episode.
For SB99, the quenching of star formation leads to a rapid decrease of $\phi$ in bins A and B (dotted/dashed lines with a red colour palette). This is not the case for BPASS, where the presence of binary stars keeps the EUV flux high, showing only some negligible differences in bin A (dotted/dashed lines with a blue colour palette). This behaviour is comparable to what has already been discussed in literature by \citet{Gotberg+19} and presented for single starburst in \citet{stanway}.

The different hardness of BPASS vs. SB99 SEDs, for varying SFHs, affects the transition from the HII region to the PDR. This is shown in Figure \ref{fig:temp}, where we plot the free-electron temperature ($T_e$), and the H$^+$ number density ($n_{H^+}$) as a function of the gas column density for two CLOUDY models: $\log U=-4.00$, $\log n=3.00$, $Z=0.55\, Z_{\odot}$ (left plots), $\log U=-2.00$, $\log n=1.00$, $Z=0.55\, Z_{\odot}$  (right plots). These are representative of typical ISM conditions inferred for ALPINE (see Table \ref{table:ALPINE_data}). The transition from the HII region to the PDR is marked by a drop in $T_e$ and $n_{H^+}$.
In the continuous SF case, the HII regions created by BPASS and SB99 approximately extend to comparable hydrogen column densities ($N_H\sim 10^{19.2}\, \rm cm^{-2}$ and $N_H\sim 10^{20.8}\, \rm cm^{-2}$, respectively) and comparable temperature ($T\sim10^{4}\, \rm K$).
The intensity of the ionising radiation determines the thickness of the HII region: an higher $U$ results in a larger HII region.
For the quenched SF cases, the BPASS SED keeps the HII region depth and temperature to values comparable to those produced by a continuous SF.
On the other hand, the drop in the ionising photon budget for SB99 models causes the HII sizes to shrink, and their temperature to decrease.
For larger $U$, the temperature drop is sharper, clearly separating the HII region from the PDR. Smaller $U$ result in a smoother temperature drop, leading to a gradual transition from the HII region to the PDR. The same holds true for the $n_{H^+}$ profile.
These differences, as we will discuss later, affect the nebular emission from the ionised gas, and the line ratios with respect to the [CII].

\subsection{CLOUDY output}\label{sec:cloudy_output}

\begin{figure*}[htbp]
    \centering
    \includegraphics[width=16cm]{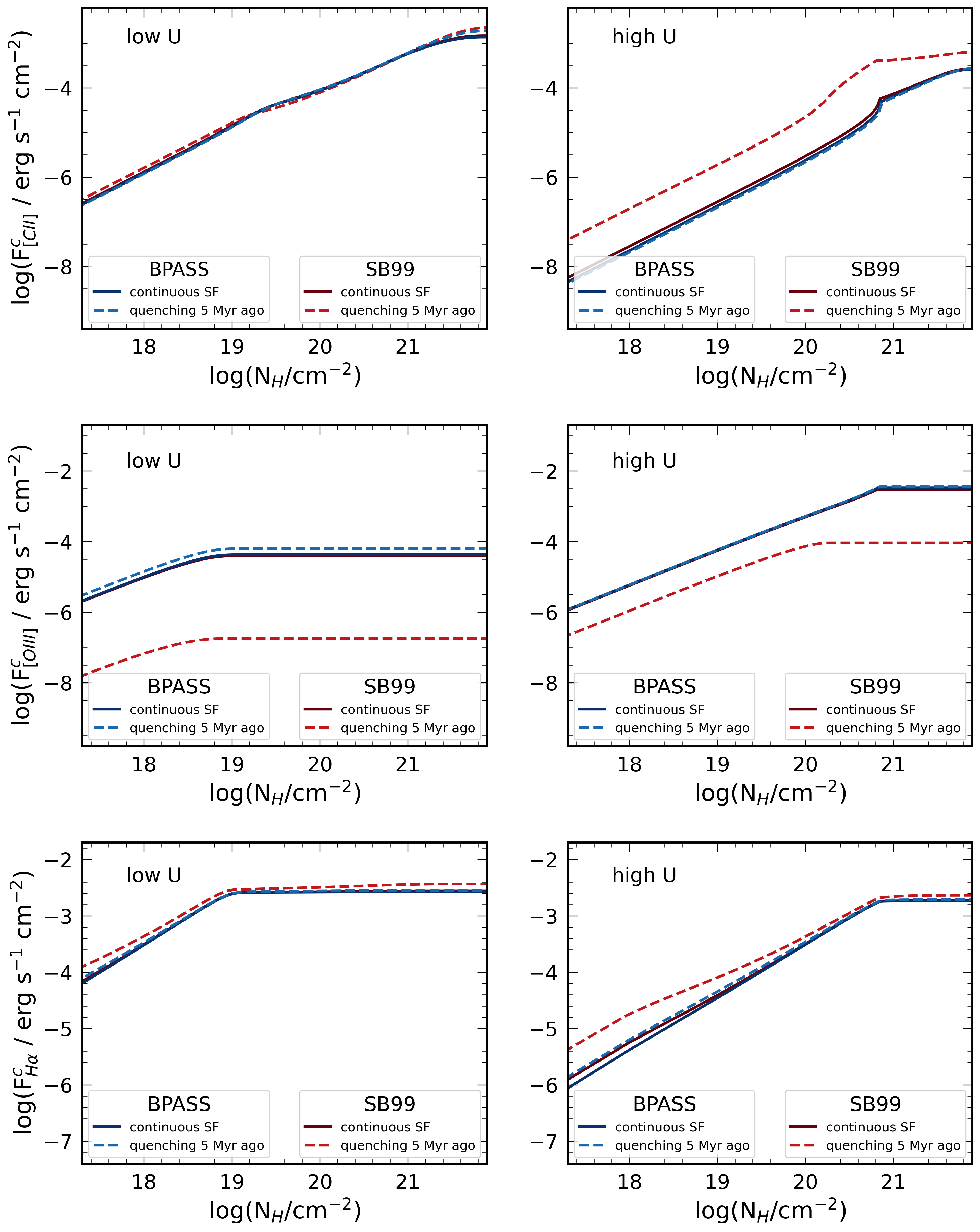}
    \caption{The cumulative flux of [CII] $158\rm \, \mu m$ (upper plots), [OIII]$5007\text{\AA}$ (middle plots) and H$\alpha$ (lower plots) as a function of the hydrogen column density ($N_H$). \textsc{CLOUDY} models $\log U=-4.00$, $\log n=3.00$, $Z=0.55\, Z_{\odot}$ (left plots, low $U$), $\log U=-2.00$, $\log n=1.00$, $Z=0.55\, Z_{\odot}$  (right plots, high $U$). The different SFHs are represented with solid (continuous SF), dashed (quenching $\Delta{t}_q=5$ Myr) lines, respectively. Red colour palette lines are associated to SB99 models, blue colour palette lines to BPASS models.}
    \label{fig:cumulative}
\end{figure*}

\begin{figure*}[htbp]
    \centering
    \includegraphics[width=16cm]{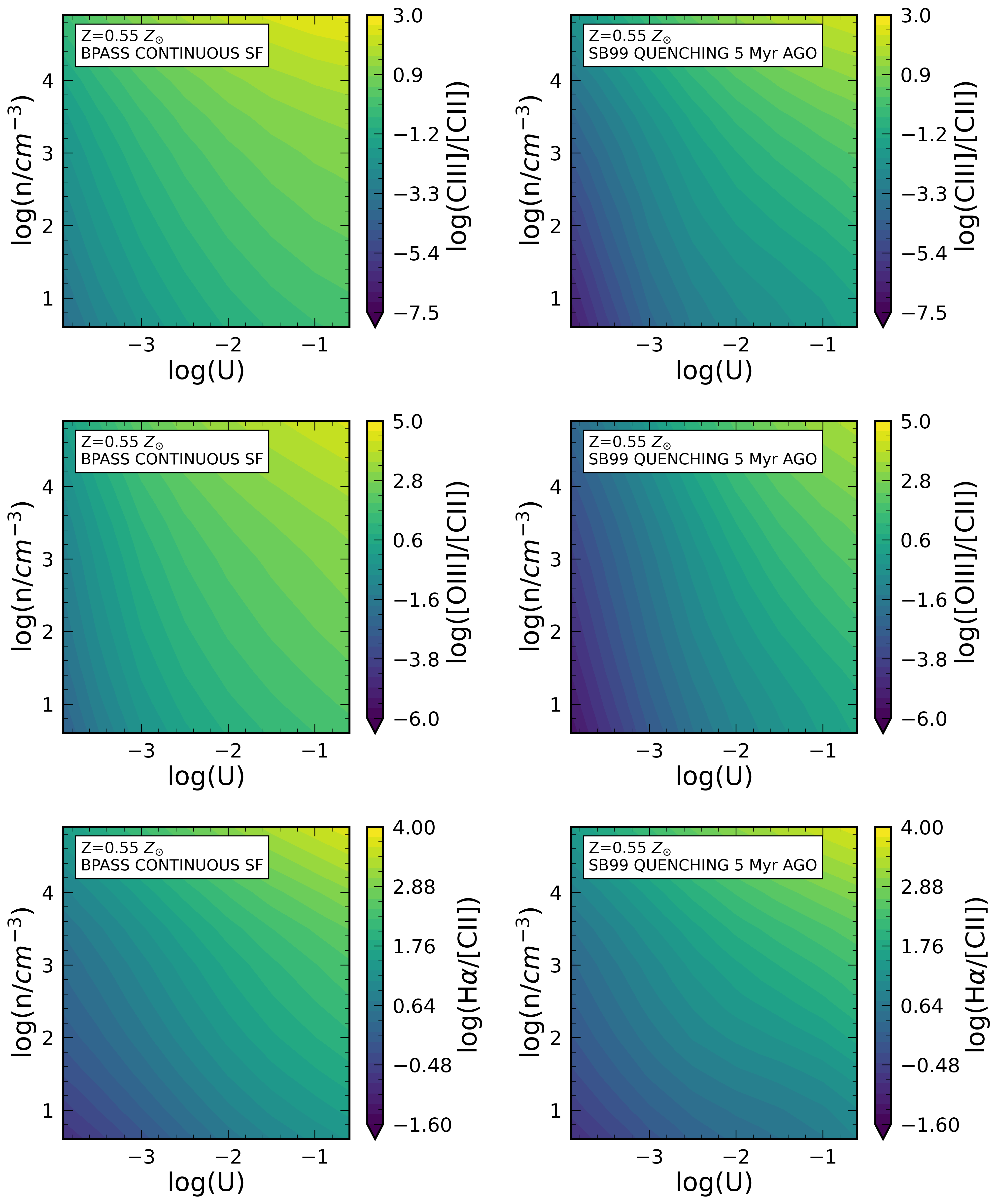}
    \caption{Contour plots of the ratio between the cumulative flux of CIII]$1909\rm \text{\AA}$ (upper panels), [OIII]$5007\rm \text{\AA}$ $\,$(middle panels), H$\alpha$ (lower panels) and the [CII] line, as a function of the ionisation parameter $U$, and the H number density $n$. The gas metallicity is fixed to $Z=0.55\, \rm Z_{\odot}$. Left panels: CLOUDY models using BPASS with continuous SFH. Right panels: CLOUDY models using SB99 with quenched SFH ($\Delta t_q = 5\, \rm Myr$).}
    \label{fig:fractions}
\end{figure*}

The \textsc{CLOUDY} outputs used in this work are the cumulative emergent fluxes ($\varepsilon_{line}(n,\,U,\, Z)$ in $\rm erg\,s^{-1}\, cm^{-2}$) of selected emission lines as a function of $N_{\rm H}$. We consider $\varepsilon_{line}$ at the end of the gas slab, $N_H=10^{22}\, \mathrm{cm^{-2}}$. This is a conservative assumption because HII region tracers reach the plateau at $N_H\sim 10^{19.2}\, \rm cm^{-2}$ and $N_H\sim 10^{20.8}\, \rm cm^{-2}$, respectively for low $U$ (left plots) and high $U$ (right plots, Figure \ref{fig:temp}). Among all possible lines, we focus on the emergent [CII] $158\, \rm \mu m$ flux, as the ratios with respect to [CII] will be used to anchor our predictions for the nebular lines (H$\beta$, [OIII] $5007\rm \text{\AA}$, H$\alpha$, [NII] $6583\rm \text{\AA}$). Moreover, we also focus on the CIII]$1909\rm \text{\AA}$ flux because, among UV tracers, theoretical studies \citep{Feltre+16,Jaskot+16,Nakajima+18} suggest that galaxies with low metallicity at $z>5$ are expected to show prominent CIII] emission. Indeed, the current highest redshift source spectroscopically confirmed with JWST \citep[$z=14.32$,][]{Carniani+24} has a tentative CIII] $1907,1909\,\rm \text{\AA}$ doublet detection. Additionally, CIII] and [CII] emissions arise from the same element, so their ratio is unaffected by possible variations in the relative elemental abundances with $Z$.
Through the paper we discuss in details the CIII] $1909 \rm \text{\AA}$, [OIII] $5007 \rm \text{\AA}$ and H$\alpha$ emission while the interested reader is referred to the Appendix \ref{sec:appendix} for predictions concerning [NII] $6583\rm \text{\AA}$ and H$\beta$.

In the upper panels of Figure \ref{fig:cumulative} we plot the [CII] cumulative flux, as a function of the hydrogen column density $N_H$, for \textsc{CLOUDY} models corresponding to $\log U=-4.00$, $\log n=3.00$, $Z=0.55\, Z_{\odot}$ (left, low $U$), $\log U=-2.00$, $\log n=1.00$, $Z=0.55\, Z_{\odot}$  (right, high $U$), but characterised by different SFHs and/or stellar population. In general, all the models show a similar [CII] profile. However, in the high $U$ case the [CII] cumulative flux produced by the SB99 quenched SF is systematically higher with respect to those produced by the other models.
This is linked to the excitation energy of the transition \citep[$T_{ex}\approx92\, \rm{K}$,][]{Ostembrook2006}, making the [CII] emission only mildly sensitive to temperature variations in the PDR above $\approx 100\,\rm K$. As can be noticed by looking at Fig. \ref{fig:temp}, the PDR plateau temperature of all the models is $T_e\sim100-200\,\rm K$, but in the SB99 quenched SF case the PDR is more extended, thus boosting the [CII] emission.

In the middle panels of Figure \ref{fig:cumulative} we plot the cumulative flux of [OIII] as a function of $N_H$ for low $U$ and high $U$.
We observe a saturation of the cumulative flux of the line at $N_H\sim 10^{19.2}\, \rm cm^{-2}$ (left) and $N_H\sim 10^{20.8}\, \rm cm^{-2}$ (right), corresponding to the end of the HII region. This mirrors the temperature profile (see Figure \ref{fig:temp}) that in the case of the SB99 quenched SFH drops at a slightly lower $N_H$ in both the low and high $U$ cases. 
The [OIII] traces the ionised gas phase; thus an increasing $U$ boosts the saturation plateau for all models.
At fixed $N_H$, we observe that in both low $U$ and high $U$ cases, the cumulative flux of [OIII] for SB99 quenched SF (red dashed lines) is systematically lower with respect to the other cases.
This is due to the softer spectrum returned by SB99 for a quenched SF that reduces the thickness and temperature of the HII region (see Section \ref{sec:SB99vsBPASS}). The [OIII] line is, indeed, extremely sensitive to temperature variations  due to the high excitation energy \citep[$T_{ex}>20000\, \rm{K}$,][]{Ostembrook2006}. Even small $T_e$ variations in the HII region, caused by variations in $U$, SFH, and/or binary presence, have a significant impact on [OIII] emission. This is true for all the UV and optical lines that have a high excitation energy.

Finally, in the bottom panels of Figure \ref{fig:cumulative} we show the cumulative flux of H$\alpha$.
At any fixed value of $N_H$, the H$\alpha$ cumulative flux in the SB99 quenched SF scenario (red dashed lines) is higher compared to the other cases.
This depends on the SED normalisation with $U$. 
In fact, at fixed $U$ the total number of photons above the Lyman limit ($h\nu>13.6\, \rm eV$) is the same, but for the (softer) SED resulting from quenched SF this translates into an increase in the number of photons near $13\, \rm eV$. This causes a slight enhancement in the H$\alpha$ emission.
Moreover, the temperature variation of the HII region minimally affects the H$\alpha$ flux \citep[emissivity ratio $\epsilon_{H\alpha}/\epsilon_{H\beta}$ at fixed density $\log n=2\, \rm{cm^{-3}}$ is $3.0$, $2.8$ and $2.7$ respectively at $T_{ex}=5000,\, 10000,\, 20000\, \rm{K}$; ][]{Ostembrook2006}.
\\

In Figure \ref{fig:fractions} we show the contour plots of the flux ratios between nebular lines and the [CII], as a function of $n$ and $U$, assuming $Z=0.55\, \rm Z_{\odot}$.
By comparing the observed nebular-to-[CII] line ratios with the theoretical contours, one can put constraints on the ISM properties characterising the targeted galaxies.

In the upper panels, we show the CIII]/[CII] ratio for BPASS continuous SF (left plot) and SB99 quenched SF (right plot).
The ratio increases with increasing $U$ and $n$.
This is due to the rise of the cloud temperature/widening of the HII region with $U$ (see Figures \ref{fig:temp}, \ref{fig:cumulative}), and to the higher CIII] critical density with respect to that of [CII].
SB99 quenched SF case shows, for fixed $U$ and $n$, a systematically lower ratio with respect to the BPASS continuous SF case. This is due, as already discussed, to the softer spectrum returned by SB99 quenched SF case.
In the middle panels, we show instead the [OIII]/[CII] ratio for BPASS continuous SF and SB99 quenched SF cases. This ratio shows a similar behaviour with respect to the CIII]/[CII] because of the same physical reasons. In the lower panels, we finally plot the H$\alpha$/[CII] ratio. As in the cases discussed above, the ratio increases with increasing $n$ and $U$.
However, the ratio remains the same between the BPASS continuous SF case and the SB99 quenched case for fixed $U$ and $n$ values.
This trend can be explained by looking at Figure \ref{fig:cumulative}.
In both [CII] and H$\alpha$ cases, the cumulative flux at $N_H=10^{22}\,\rm cm^{-2}$ increases in the SB99 quenched SF case, resulting in a ratio that remains the same as in the BPASS continuous SF case.\\
The trends discussed so far are fundamental to interpret the line luminosities prediction for ALPINE galaxies that we will discuss in Section \ref{sec:predictions}.

\section{Model Application}\label{sec:Model_appl}

\subsection{Dataset}\label{sec:dataset}

ALPINE \citep[][]{Bethermin,Fevre,Faisst} is an ALMA large programme aimed at measuring [CII] $158\, \mathrm{\mu m}$ and rest-frame FIR continuum emission from a representative sample of 118 main-sequence galaxies at $4.4<z<5.8$. Observations
were carried out between May 2018 and February 2019 \citep[][]{Bethermin,Faisst}.
All galaxies are located in the Cosmic Evolution Survey \citep[COSMOS,][]{Scoville+07}, and in the Extended Chandra Deep Field South \citep[ECDFS,][]{Giacconi+02} fields.
The sample was selected to exclude as much as possible galaxies with evident AGN signature. However, a low luminosity AGN was recently discovered in one source of the sample \citep[GDS J033218.92-275302.7, ][]{Barchiesi+23,Ubler+23}.
ALPINE galaxies are characterised by SFR$\gtrsim 10\, \mathrm{M_{\odot}\,yr^{-1}}$ and stellar masses $M_{\star}\sim 10^9-10^{11}\, \mathrm{M_{\odot}}$ \citep[][]{Faisst}.
Among the 118 sources, 75 and 23 are \rm{[CII]} and $158\, \rm \mu m$ continuum detected, respectively.

\begin{figure}[htbp]
    \centering
    \resizebox{\hsize}{!}{\includegraphics{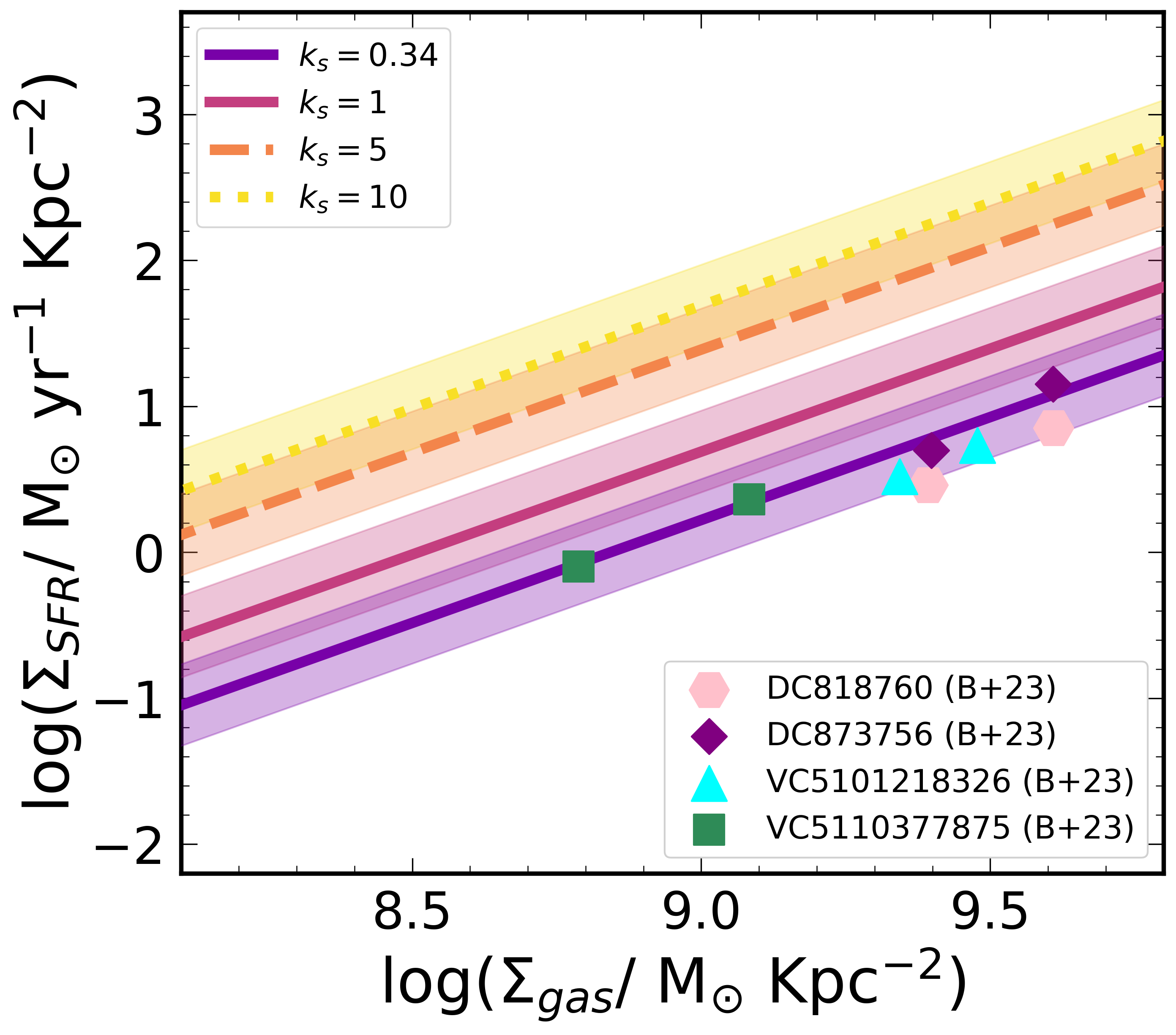}}
    \caption{The Kennicutt-Schmidt law \citep[solid magenta line, ][]{Heiderman+10}. The coloured points represent the location on the KS plane of the subset of ALPINE galaxies analysed by \citet{Bethermin24}.
    Each galaxy is resolved in two different beams, providing two measures for $\Sigma_{\rm gas}$ and $\Sigma_{\rm SFR}$.
    Following the notation of \cite{Ferrara} this corresponds to a burstiness parameter $k_s=0.34$ (purple line). For comparison, we also report the relation obtained substituting $k_s=5,\, 10$ in Equation (\ref{eq:ks}), with orange dashed and yellow dotted lines, respectively.
    The y-dispersion is $\sigma=0.28$.}
    \label{fig:KS}
\end{figure}
For the present analysis, we select a subset of 44 galaxies with measured \rm{[CII]} luminosity \citep[][]{Faisst} and UV radius \citep[][]{Fujimoto}.
These galaxies represent the complete subsample of \citet{Fujimoto}, excluding two galaxies lacking in SFR and UV radius data (Table \ref{table:ALPINE_data}).
The UV radius data, together with the SFR, are needed to exploit the analytical model described in Section \ref{sec:deriv_pred} used to infer the fiducial $n$ and $U$ values for the ALPINE galaxies. In our work we consider the total SFR, SFR$_{\rm UV+IR}$, namely the sum of the unobscured (SFR$_{\rm UV}$) and obscured (SFR$_{\rm IR}$) one. We derive SFR$_{\rm UV}$ from the UV luminosity at $1500\rm \text{\AA}$ \citep[][]{Faisst} uncorrected for dust extinction, and SFR$_{\rm IR}$ from the IR luminosity reported in \citep[][]{Bethermin+20}, using the \citet{Kennicutt} relation. For those galaxies that are not detected in continuum with ALMA, we derive the $L_{\rm IR}$ from the infrared excess (IRX=$L_{\rm IR}/L_{\rm UV}$) computed for the ALPINE galaxies by \citet{Fudamoto+20} from the ultraviolet spectral slope $\beta$.

\subsection{Fiducial CLOUDY models for ALPINE}\label{sec:deriv_pred}

Following \citet{Ferrara} (F19 hereafter), we assume that each galaxy can be represented as a flat, parallel slab of gas with an uniform number density $n$, metallicity $Z$, and total gas column density $\Sigma_{\rm gas}$. The slab is illuminated by a (stellar) source of ionising and non-ionising radiation. F19 allows us to infer $(n,\, U)$ -- that are two of the three free parameters of the \textsc{CLOUDY} models (see Sec. \ref{sec:CLOUDY}) -- by linking them to global galaxy properties by using the following relations:
\begin{eqnarray}\label{eq:SAM}
    n&=&5.4 \times 10^{-13} {\Sigma_{\rm gas}}^2 {\sigma^{-2}}\, [\rm{cm^{-3}}], \\
    U&=&1.7 \times 10^{14} \frac{\Sigma_{\rm SFR}}{{\Sigma_{\rm gas}}^2}
\end{eqnarray} 

\begin{table*}[htbp]
\caption{Physical properties of the 44 ALPINE galaxies considered in this work.}
\label{table:ALPINE_data}
\resizebox{\textwidth}{!}{
\begin{tabular}{ccccccccc}
\toprule
\toprule
\multirow{2}{*}{Galaxy ID} & \multirow{2}{*}{$\rm z_{[CII]}$} & \multirow{2}{*}{$\rm log\left(\frac{SFR_{tot}}{M_{\odot}\,yr^{-1}}\right)$} & \multirow{2}{*}{$\rm log\left(\frac{SFR_{tot}^ {up}}{M_{\odot}\,yr^{-1}}\right)$} & \multirow{2}{*}{$\rm log\left(\frac{SFR_{tot}^ {low}}{M_{\odot}\,yr^{-1}}\right)$} & \multicolumn{4}{c}{(log U, log n)}                                 \\
                           &                                                                  &     &                                                                   &                                                                          & $k_s=0.34$ "fiducial" & $k_s=1$      & $k_s=5$      & $k_s=10$     \\
\midrule
C\_GOODSS\_32 *              & 4.4                & 1.74                 & 1.83                & 1.62                  & (-3.8, 2.7)           & (-3.1, 2.1)           & (-2.1, 1.1)           & (-1.7, 0.6)           \\
DC\_308643                 & 4.5                & 1.72                 & 1.77                & 1.67                  & (-3.8, 2.7)           & (-3.1, 2.1)           & (-2.1, 1.1)           & (-1.7, 0.6)           \\
DC\_351640                 & 5.7                & 1.23                 & 1.43                & 1.03                  & (-4.0, 3.7)           & (-3.4, 3.0)           & (-2.4, 2.0)           & (-1.9, 1.6)           \\
DC\_372292                 & 5.1                & 1.59                 & 1.65                & 1.51                  & (-3.7, 2.4)           & (-3.0, 1.7)           & (-2.0, 0.7)           & (-1.6, 0.3)           \\
DC\_396844 *                 & 4.5                & 1.93                 & 2.00                & 1.84                  & (-4.0, 3.6)           & (-3.3, 2.9)           & (-2.3, 1.9)           & (-1.9, 1.5)           \\
DC\_416105                 & 5.6                & 1.28                 & 1.43                & 1.14                  & (-3.5, 2.0)           & (-2.9, 1.3)           & (-1.9, 0.3)           & (-1.4, -0.1)          \\
DC\_417567 *                 & 5.7                & 2.00                 & 2.08                & 1.91                  & (-4.0, 3.5)           & (-3.3, 2.9)           & (-2.3, 1.9)           & (-1.9, 1.4)           \\
DC\_422677 *                 & 4.4                & 1.94                 & 2.05                & 1.79                  & (-4.0, 3.6)           & (-3.3, 2.9)           & (-2.3, 1.9)           & (-1.9, 1.5)           \\
DC\_432340                 & 4.4                & 1.73                 & 1.78                & 1.69                  & (-3.7, 2.6)           & (-3.1, 2.0)           & (-2.1, 1.0)           & (-1.6, 0.5)           \\
DC\_434239                 & 4.5                & 1.75                 & 1.82                & 1.68                  & (-3.5, 1.8)           & (-2.8, 1.2)           & (-1.8, 0.2)           & (-1.4, -0.3)          \\
DC\_454608                 & 4.6                & 1.56                 & 1.60                & 1.52                  & (-3.7, 2.6)           & (-3.0, 1.9)           & (-2.0, 0.9)           & (-1.6, 0.5)           \\
DC\_488399 *                 & 5.7                & 1.95                 & 2.01                & 1.87                  & (-4.1, 3.9)           & (-3.4, 3.2)           & (-2.4, 2.2)           & (-2.0, 1.8)           \\
DC\_493583 *                 & 4.5                & 1.79                 & 1.90                & 1.65                  & (-3.9, 3.3)           & (-3.2, 2.6)           & (-2.2, 1.6)           & (-1.8, 1.2)           \\
DC\_494057 *                 & 5.5                & 1.87                 & 1.95                & 1.78                  & (-4.0, 3.5)           & (-3.3, 2.8)           & (-2.3, 1.8)           & (-1.9, 1.4)           \\
DC\_494763                 & 5.2                & 1.20                 & 1.41                & 1.03                  & (-3.7, 2.4)           & (-3.0, 1.7)           & (-2.0, 0.7)           & (-1.6, 0.3)           \\
DC\_519281                 & 5.6                & 1.38                 & 1.46                & 1.30                  & (-3.8, 3.0)           & (-3.2, 2.3)           & (-2.2, 1.3)           & (-1.7, 0.9)           \\
DC\_536534                 & 5.7                & 1.59                 & 1.69                & 1.48                  & (-3.7, 2.6)           & (-3.0, 1.9)           & (-2.0, 0.9)           & (-1.6, 0.5)           \\
DC\_539609 *                 & 5.2                & 1.85                 & 1.93                & 1.74                  & (-3.9, 3.1)           & (-3.2, 2.4)           & (-2.2, 1.4)           & (-1.8, 1.0)           \\
DC\_552206 *                 & 5.5                & 2.04                 & 2.14                & 1.93                  & (-3.9, 3.1)           & (-3.2, 2.4)           & (-2.2, 1.4)           & (-1.8, 1.0)           \\
DC\_627939                 & 4.5                & 1.66                 & 1.72                & 1.59                  & (-3.7, 2.7)           & (-3.1, 2.0)           & (-2.1, 1.0)           & (-1.7, 0.6)           \\
DC\_630594                 & 4.4                & 1.48                 & 1.54                & 1.42                  & (-3.7, 2.6)           & (-3.0, 1.9)           & (-2.0, 0.9)           & (-1.6, 0.5)           \\
DC\_683613 *                 & 5.5                & 1.88                 & 1.97                & 1.77                  & (-4.0, 3.5)           & (-3.3, 2.9)           & (-2.3, 1.9)           & (-1.9, 1.4)           \\
DC\_709575                 & 4.4                & 1.45                 & 1.48                & 1.40                  & (-3.7, 2.6)           & (-3.1, 2.0)           & (-2.1, 1.0)           & (-1.6, 0.5)           \\
DC\_733857                 & 4.5                & 1.61                 & 1.66                & 2.30                  & (-3.9, 3.1)           & (-3.2, 2.4)           & (-2.2, 1.4)           & (-1.8, 1.0)           \\
DC\_773957                 & 5.7                & 1.24                 & 1.45                & 3.26                  & (-3.6, 2.3)           & (-2.9, 1.6)           & (-1.9, 0.6)           & (-1.5, 0.2)           \\
DC\_818760 *                 & 4.6                & 2.38                 & 2.43                & 2.33                  & (-4.1, 3.9)           & (-3.4, 3.2)           & (-2.4, 2.2)           & (-2.0, 1.8)           \\
DC\_834764                 & 4.5                & 1.49                 & 1.58                & 1.41                  & (-3.7, 2.4)           & (-3.0, 1.8)           & (-2.0, 0.8)           & (-1.6, 0.3)           \\
DC\_842313                 & 4.6                & 2.20                 & 2.25                & 2.15                  & (-3.7, 2.5)           & (-3.0, 1.8)           & (-2.0, 0.8)           & (-1.6, 0.4)           \\
DC\_848185 *                 & 5.3                & 2.10                 & 2.15                & 2.04                  & (-3.9, 3.3)           & (-3.2, 2.6)           & (-2.3, 1.6)           & (-1.8, 1.2)           \\
DC\_873321                 & 5.2                & 1.78                 & 1.85                & 1.71                  & (-3.8, 2.8)           & (-3.1, 2.1)           & (-2.1, 1.1)           & (-1.7, 0.7)           \\
DC\_873756 *                 & 4.5                & 2.45                 & 2.47                & 2.42                  & (-4.0, 3.5)           & (-3.3, 2.9)           & (-2.3, 1.9)           & (-1.9, 1.4)           \\
DC\_880016                 & 4.5                & 1.39                 & 1.47                & 1.31                  & (-3.7, 2.5)           & (-3.0, 1.8)           & (-2.0, 0.8)           & (-1.6, 0.4)           \\
DC\_881725 *                 & 4.6                & 1.92                 & 2.02                & 1.80                  & (-4.0, 3.4)           & (-3.3, 2.7)           & (-2.3, 1.7)           & (-1.9, 1.3)           \\
vc\_5100537582             & 4.6                & 1.13                 & 1.25                & 1.01                  & (-3.7, 2.5)           & (-3.0, 1.9)           & (-2.0, 0.9)           & (-1.6, 0.4)           \\
vc\_5100541407             & 4.6                & 1.50                 & 1.57                & 1.43                  & (-3.7, 2.5)           & (-3.0, 1.8)           & (-2.0, 0.8)           & (-1.6, 0.4)           \\
vc\_5100559223             & 4.6                & 1.44                 & 1.52                & 1.36                  & (-3.6, 2.2)           & (-2.9, 1.6)           & (-1.9, 0.6)           & (-1.5, 0.1)           \\
vc\_5100822662 *             & 4.5                & 1.78                 & 1.84                & 1.70                  & (-3.6, 2.3)           & (-3.0, 1.7)           & (-2.0, 0.7)           & (-1.5, 0.2)           \\
vc\_5100969402 *             & 4.6                & 1.90                 & 2.00                & 1.76                  & (-4.0, 3.5)           & (-3.3, 2.8)           & (-2.3, 1.8)           & (-1.9, 1.4)           \\
vc\_5100994794 *             & 4.6                & 1.54                 & 1.64                & 1.40                  & (-3.5, 1.7)           & (-2.8, 1.1)           & (-1.8, 0.1)           & (-1.4, -0.4)          \\
vc\_5101218326 *             & 4.6                & 2.08                 & 2.14                & 2.01                  & (-3.7, 2.6)           & (-3.1, 2.0)           & (-2.1, 1.0)           & (-1.6, 0.6)           \\
vc\_510786441              & 4.5                & 1.71                 & 1.78                & 1.64                  & (-3.8, 2.8)           & (-3.1, 2.1)           & (-2.1, 1.1)           & (-1.7, 0.7)           \\
vc\_5110377875             & 4.6                & 1.91                 & 1.98                & 1.84                  & (-3.8, 3.0)           & (-3.2, 2.3)           & (-2.2, 1.3)           & (-1.7, 0.9)           \\
vc\_5180966608 *             & 4.5                & 2.00                 & 2.07                & 1.91                  & (-4.0, 3.4)           & (-3.3, 2.8)           & (-2.3, 1.8)           & (-1.9, 1.3)           \\
v\_efdcs\_530029038 *        & 4.4                & 1.64                 & 1.74                & 1.50                  & (-3.4, 1.7)           & (-2.8, 1.0)           & (-1.8, 0.0)           & (-1.3, -0.4)          \\
\midrule
average                    & 4.8                & 1.72                 & --                  & --                    & (-3.8, 2.9)           & (-3.1, 2.2)           & (-2.1, 1.2)           & (-1.7, 0.8)          \\
\bottomrule
\end{tabular}
}
\parbox{\textwidth}{\tablefoot{The ionisation parameter $U$ and the hydrogen density $n$ have been estimated for different values of the burstiness parameter $k_s$ (see Sec. \ref{sec:deriv_pred}).
SFR$_{tot}^{up}$ and SFR$_{tot}^{low}$ are the upper and lower bounds of the total SFR.\\
\tablefoottext{*}{galaxies infrared detected}}}
\end{table*}

where $\Sigma_{\rm gas}$ is the gas surface density, $\Sigma_{\rm SFR}$ is the SFR surface density, and $\sigma$ is the rms turbulent velocity in $\rm km/s$.  We derive $\Sigma_{\rm SFR}$ from the SFR$_{\rm UV+IR}$ and the IR radius ($r_{\rm IR}$) for the ALPINE galaxies.
Given the recent results by \citet{Pozzi+24} in ALPINE and \citet{Mitsuhashi} in CRISTAL, we assume $r_{\rm IR}=2 r_{\rm UV}$.
We assume $\sigma\approx 50 \, \rm{km \, s^{-1}}$, the average value found for the ALPINE sample by \citet{Jones+21,Parlanti+23} with low resolution observations.
This value approaches the upper limits measured in high-resolution studies of starburst galaxies \citep[$\sigma \approx 40 \, \rm{km \, s^{-1}}$][]{Rizzo+21,Lelli+21,Roman-Olivera+23}.
Choosing a reasonable $\sigma$ is crucial due to its impact in estimating the gas density, $n$. Halving $\sigma$ quadruples the value of $n$, as shown in Eq. (\ref{eq:SAM}).

We compute instead $\Sigma_{\rm gas}$ using the Kennicutt-Schmidt (KS) relation \citep{Heiderman+10} adopting the notation introduced by F19:
\begin{equation}\label{eq:ks}
    \Sigma_{\rm SFR}=10^{-12} k_s  \Sigma_{\rm gas}^m \;\;[\rm{M_{\odot}\, yr^{-1}\, kpc^{-2}}]\;\; (m=1.4).
\end{equation}
In Equation (\ref{eq:ks}), F19 defined the burstiness parameter ($k_s$), defined as the deviation from the KS relation.
Galaxies with $k_s>1$ have a larger SFR per unit area
with respect to those located on the KS relation, so they are starburst and can convert gas in stars more efficiently with a shorter depletion time ($t_{dep}$). 
\begin{table*}[htbp]
\centering
\caption{The slope $\alpha$ and the intercept $\beta$ of the linear regression for the CIII] $1909\rm \text{\AA}$, [OIII] $5007\rm \text{\AA}$ and H$\alpha$ lines.}       
\label{table:RESULTS_1}
\resizebox{0.9\textwidth}{!}{
\begin{tabular}{cccccccc}
\toprule
\toprule
\multicolumn{1}{c}{\multirow{2}{*}{SFH}} & \multicolumn{1}{c}{\multirow{2}{*}{$k_s$}} & \multicolumn{2}{c}{CIII{]} $1909\rm \text{\AA}$} & \multicolumn{2}{c}{{[}OIII{]} $5007\rm \text{\AA}$} & \multicolumn{2}{c}{H$\alpha$}\\
 & & $\alpha$ & $\beta$ & $\alpha$ & $\beta$ & $\alpha$ & $\beta$\\
\midrule                               
\multicolumn{1}{c}{\multirow{4}{*}{continous SF}}    & 0.34 "fiducial"     & $0.24_{-0.21}^{0.21}$ & $40.66_{-0.38}^{0.38}$ & $0.36_{-0.21}^{0.21}$ & $40.80_{-0.37}^{0.38}$ & $0.85_{-0.22}^{0.21}$ & $41.34_{-0.38}^{0.39}$ \\[0.08cm]
                                 & 1                   & $0.70_{-0.21}^{0.21}$ & $39.88_{-0.38}^{0.38}$ & $0.51_{-0.21}^{0.21}$ & $41.77_{-0.38}^{0.38}$ & $0.84_{-0.22}^{0.21}$ & $41.57_{-0.39}^{0.38}$ \\[0.08cm]
                                 & 5                   & $0.66_{-0.21}^{0.21}$ & $40.69_{-0.37}^{0.38}$ & $0.87_{-0.22}^{0.22}$ & $42.00_{-0.38}^{0.38}$ & $0.90_{-0.22}^{0.21}$ & $41.71_{-0.38}^{0.38}$ \\[0.08cm]
                                 & 10                  & $0.87_{-0.21}^{0.21}$ & $40.34_{-0.38}^{0.38}$ & $0.88_{-0.21}^{0.21}$ & $42.19_{-0.38}^{0.37}$ & $0.96_{-0.21}^{0.21}$ & $41.84_{-0.39}^{0.38}$ \\[0.08cm]
\midrule
\multicolumn{1}{c}{\multirow{4}{*}{quenching 5 Myr}} & 0.34 "fiducial"     & $0.57_{-0.21}^{0.21}$ & $39.47_{-0.37}^{0.37}$ & $0.36_{-0.21}^{0.21}$ & $40.84_{-0.38}^{0.37}$ & $0.83_{-0.21}^{0.21}$ & $41.27_{-0.37}^{0.38}$ \\[0.08cm]
                                 & 1                   & $0.73_{-0.22}^{0.21}$ & $39.84_{-0.37}^{0.39}$ & $0.52_{-0.22}^{0.21}$ & $41.78_{-0.38}^{0.37}$ & $0.81_{-0.21}^{0.21}$ & $41.56_{-0.38}^{0.38}$ \\[0.08cm]
                                 & 5                   & $0.66_{-0.21}^{0.21}$ & $40.72_{-0.38}^{0.37}$ & $0.86_{-0.21}^{0.21}$ & $42.02_{-0.38}^{0.38}$ & $0.94_{-0.21}^{0.21}$ & $41.65_{-0.37}^{0.37}$ \\[0.08cm]
                                 & 10                  & $0.87_{-0.22}^{0.21}$ & $40.37_{-0.38}^{0.38}$ & $0.85_{-0.21}^{0.21}$ & $42.24_{-0.38}^{0.37}$ & $0.91_{-0.22}^{0.21}$ & $41.81_{-0.38}^{0.38}$\\[0.08cm]
\bottomrule
\end{tabular}
}
\parbox{\textwidth}{\tablefoot{Linear regression $\log(L_{line}/\rm erg\, s^{-1})=\alpha\log(\rm SFR/M_{\odot}\, yr^{-1})+\beta$ computed with a Monte Carlo fitting technique. The $\alpha$ and $\beta$ were obtained assuming the fiducial BPASS incident spectrum.}}
\end{table*}

Equation (\ref{eq:ks}) ultimately allows us to derive $\Sigma_{\rm gas}$ from the measured $\Sigma_{\rm SFR}$ in ALPINE as a function of the free parameter $k_s$. We assume 4 different values for the $k_s$ parameters. From the spatially resolved analysis presented by \citet{Bethermin24}  for 4 ALPINE galaxies the average value of the burstiness parameter is $0.34$ (see Figure \ref{fig:KS}) derived by converting the depletion time $t_{dep}=\Sigma_{\rm gas}/\Sigma_{SFR}\propto k_s^{-1}\Sigma_{\rm gas}^{-0.4}$. Moreover, we also consider $k_s=5$ and $k_s=10$, which are values representative of moderate and extreme starburst galaxies at high-$z$ \citep{Vallini+21}, and the “standard” KS law, namely $k_s=1$.
The galaxies presented in \citet{Bethermin24} are among the brightest [CII] galaxies in the ALPINE sample. This is why, in order to avoid bias, we include predictions for various $k_s$ values.
For $k_s=0.34$, using Eqs. (3),(4) and (5), we obtain a median value $\log U=-3.8\pm 0.2$ and $\log (n/\rm cm^{-3})=2.9\pm 0.6$ within our ALPINE subsample.
The associated errors are the subsample standard deviations for the two parameters. 
These are comparable with those inferred by \citet{Vanderhoof+22} for ten ALPINE galaxies, where the measured dust-corrected luminosity ratios of $\rm log(L_{\rm [OII]}/L_{\rm [CII]})$ and $\rm log(L_{\rm [OII]}/L_{\rm H\alpha})$ required ionisation parameters $\rm log(U)<-2$ and electron densities $\rm log(n_e/[cm^{-3}])\approx 2.5-3$. In Table \ref{table:ALPINE_data} we report the ALPINE fiducial $U$ and $n$ values and those computed for $k_s=1,\,5,\,10$. 

The inferred $n$ and $U$ for each ALPINE galaxy were then exploited to compute
\begin{equation}\label{eq:predictions_formula}
    L^{pred}_{line}=R_{line}(n,\,U,\,Z)\times L^{obs}_{\rm [CII]}
\end{equation}
from the corresponding \textsc{CLOUDY} models where:
\begin{equation}\label{eq:predictions_formula_2}
    R_{line}(n,\,U,\,Z)=\frac{\varepsilon_{line}(n,\,U,\,Z)}{\varepsilon_{\rm [CII]}(n,\,U,\,Z)}
\end{equation} 
is the CLOUDY predicted ratio of each line with respect to the [CII].  For all the ALPINE sample we assume $Z=0.5\, \mathrm{Z_{\odot}}$, similar to what has been found in \citet{Vanderhoof+22} from UV absorption lines in ALPINE sample.
The gas metallicity has a smaller impact on the line ratios with respect to $n$ and $U$. For fixed values of $n$ and $U$, the line ratio increases of $0.3-0.4$ dex when metallicity varies from $0.15\,\rm Z_{\odot}$ to $0.55\,\rm Z_{\odot}$. However, we defer a dedicated analysis to a future work.

\section{Results}\label{sec:Results}

\subsection*{Nebular line predictions and their relation with the SFR}\label{sec:predictions}
\begin{figure*}[htbp]
    \centering
    \includegraphics[width=8cm]{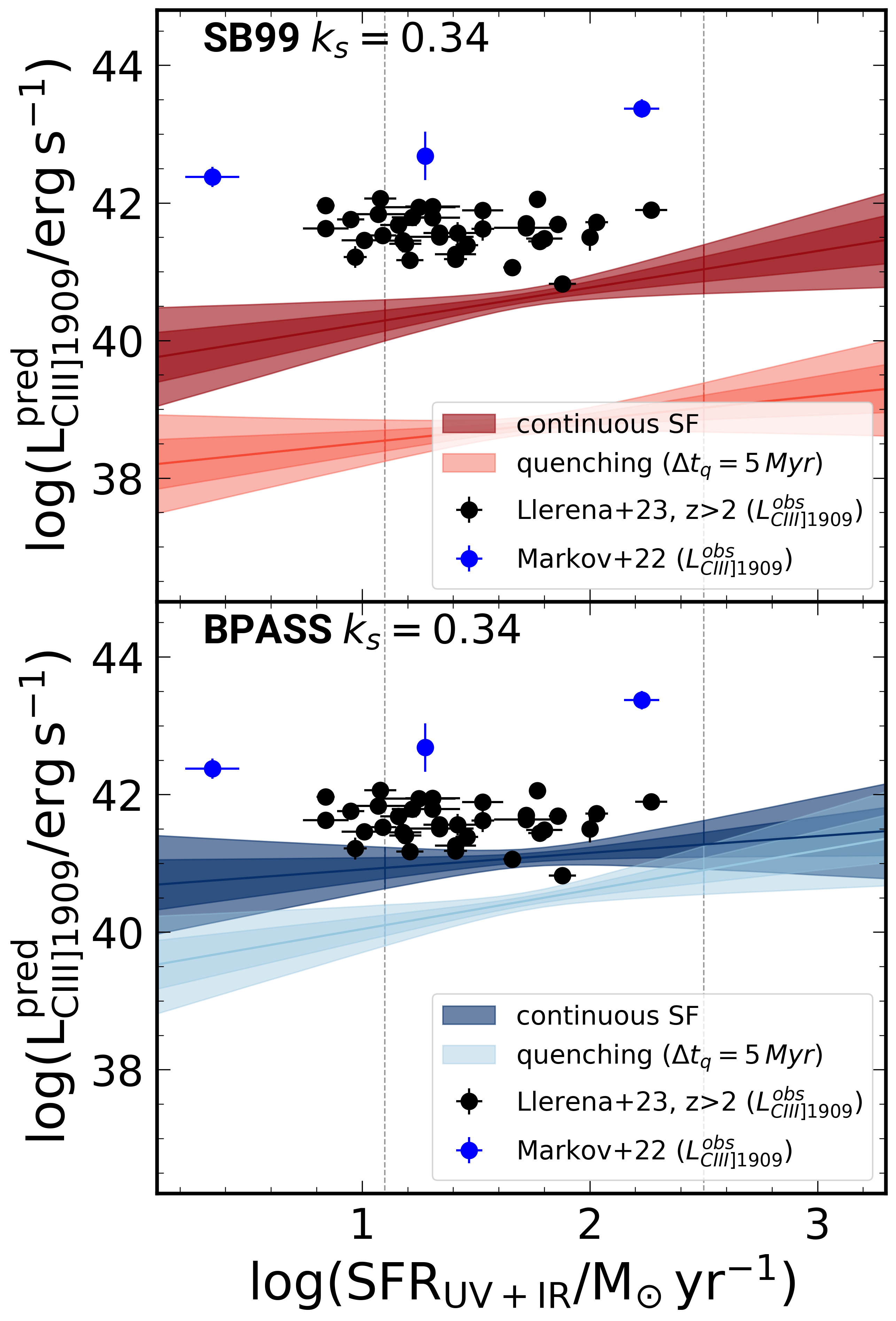}
    \includegraphics[width=8cm]{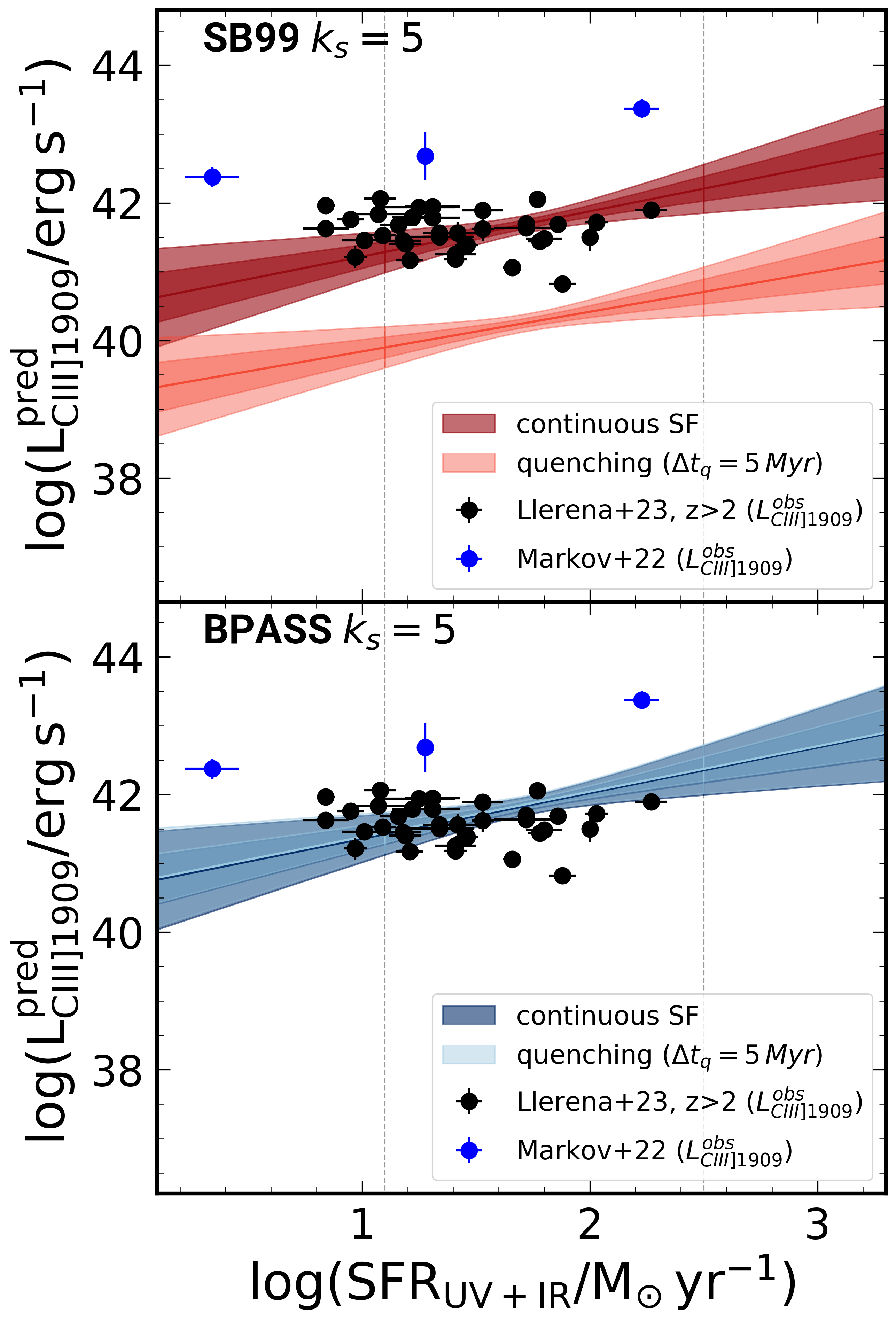}
    \caption{Predicted luminosity of CIII] $1909\rm \text{\AA}$ as a function of the total SFR (SFR$_{\rm UV+IR}$), for $k_s=0.34$ (left panel) and $k_s=5$ (right panel). The colours correspond to 1-2$\sigma$ dispersion. The central region between the grey dashed lines represents the SFR range of the ALPINE subsample modelled in this work. The red and blue palettes represent the two star formation histories considered in this work: continuous and $\Delta t_q=5\, \rm Myr$.
    The black points are the observed data from \citet{Llerena23}, the blue points from \citet{Markov}.}
    \label{fig:CIII]}
\end{figure*}

\begin{figure*}[htbp]
    \centering
    \includegraphics[width=8cm]{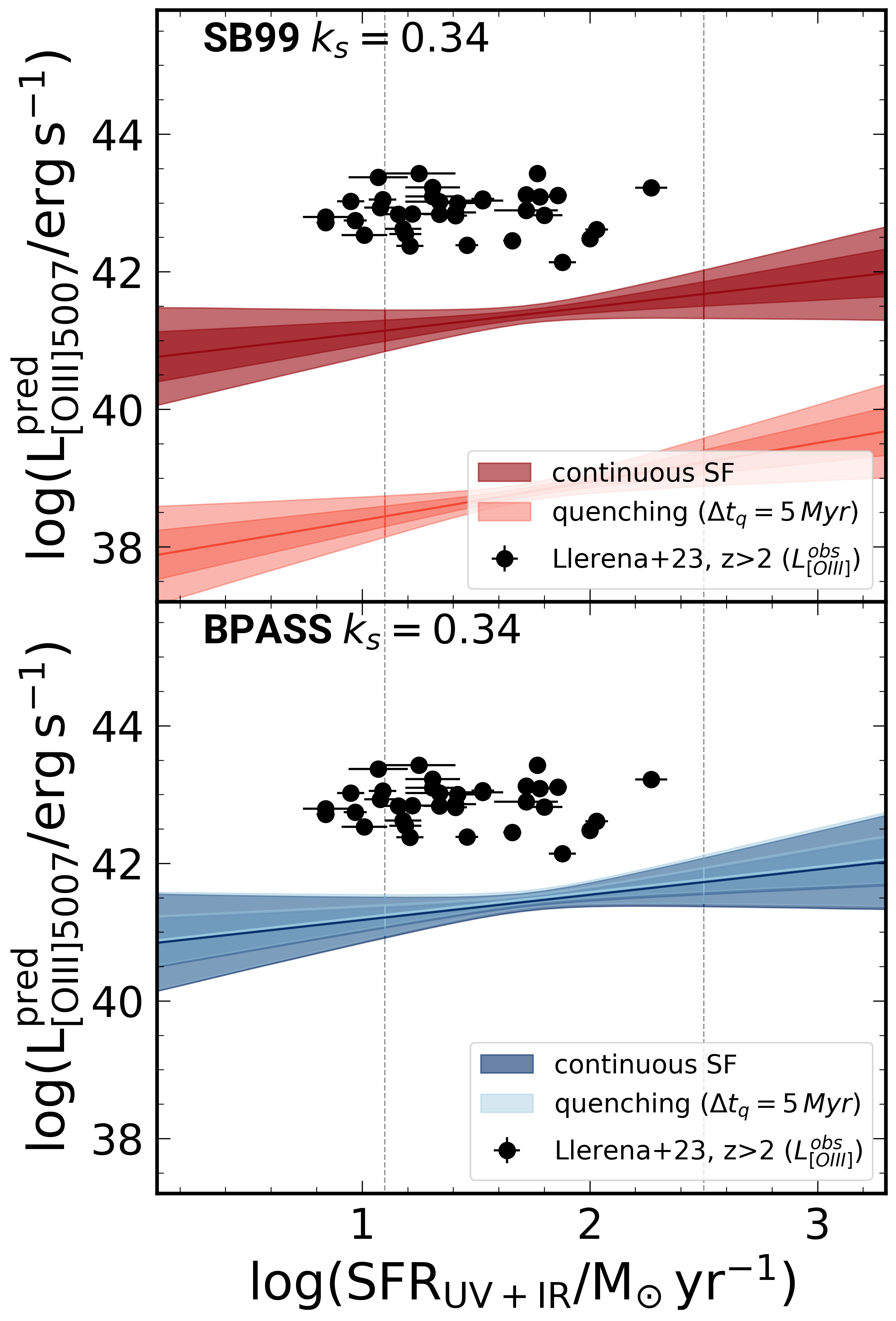}
    \includegraphics[width=8cm]{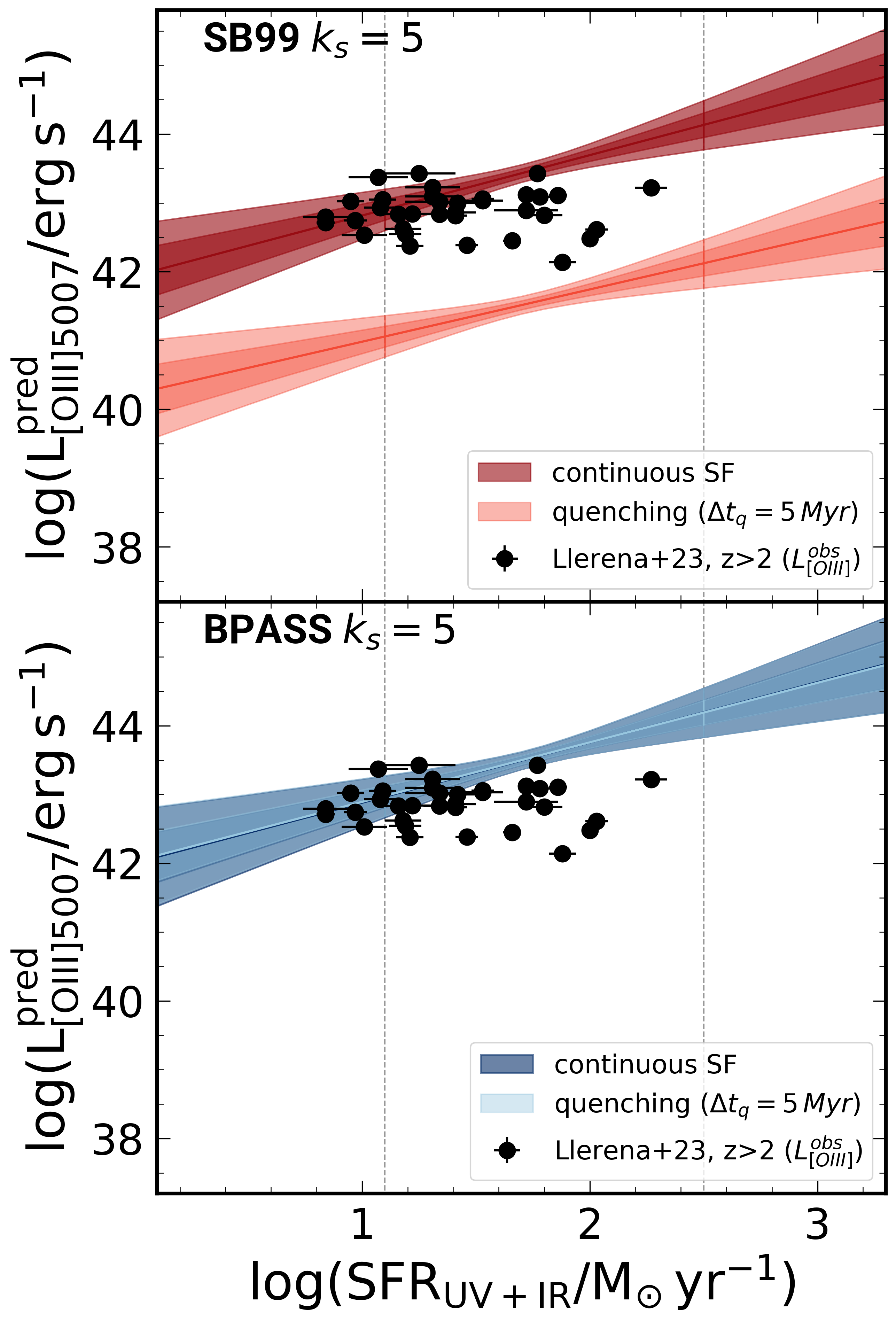}    
    \caption{Predicted luminosity of [OIII] $5007\rm \text{\AA}$ as a function of the total SFR (SFR$_{\rm UV+IR}$), for $k_s=0.34$ (left panel) and $k_s=5$ (right panel). The colours correspond to 1-2$\sigma$ dispersion. The central region between the grey dashed lines represents the SFR range of the ALPINE subsample modelled in this work. The red and blue colour palettes represent the two star formation histories considered in this work: continuous and $\Delta t_q=5\, \rm Myr$.
    The black points are the observed data from \citet{Llerena23}.}
    \label{fig:[OIII]}
\end{figure*}
Following the procedure described in Section \ref{sec:Model_appl}, for each ALPINE galaxy we derive theoretical predictions for the nebular line luminosities starting from $R_{line}(n,\,U,\,Z)$ and the observed $L_{\rm [CII]}$. We then derive relations between the theoretical line luminosity, and the total observed SFR$_{\rm UV + IR}$:
\begin{equation}
    \log(L_{line}/\rm erg\, s^{-1})=\alpha\log(\rm SFR/M_{\odot}\, yr^{-1})+\beta,
\end{equation}
by employing a Bayesian Markov Chain Monte Carlo (MCMC) fitting method implemented with the open-source Python package \texttt{emcee} \citep{emcee}. We run \texttt{emcee} with 44 random walkers exploring the parameter space for $5\times 10^{3}$ chain steps. Table \ref{table:RESULTS_1} summarises the slope ($\alpha$) and intercept ($\beta$) coefficients for the BPASS cases together with their 1$\sigma$ dispersion for the fiducial ALPINE model ($k_s=0.34$) and for $k_s=1,\,5,\,10$.\footnote{The SB99 cases are reported in Appendix (see Appendix \ref{sec:sb99predict3l}).}\\

In Figure \ref{fig:CIII]} (left panel) we show the best-fit $L_{\text{CIII]}}-$SFR$_{\rm UV+IR}$ relations, together with their 1-2$\sigma$ dispersion, for the two considered SFH and assuming $k_s=0.34$.
Figure \ref{fig:CIII]} highlights the strong impact of considering binary stars when computing the radiation illuminating the ISM. For models without binaries (SB99 cases), the CIII] is sensitive to SF quenching. In SB99 quenched SF case (upper plot), the CIII] emission drops by almost 1 dex at fixed SFR already for $\Delta{t}_q=5$ Myr. This is a consequence of the very high ionising potential of $C^+$ ($24.45\, \rm{eV}$) and of the shrinking of HII region due to a softer incident spectrum (see Sections \ref{sec:SB99vsBPASS} and \ref{sec:cloudy_output}). Assuming BPASS models instead (lower plot), these are not influenced by the presence of a quenching, as the emission remains stable up $\Delta{t}_q=5$ Myr.
This is due to binaries being able to delay the drop of the UV photon budget for a stellar population that is experiencing a quenching in SF.

In Figure \ref{fig:CIII]} (right panel) we show the best-fit L$_{\text{CIII]}}-$SFR$_{\mathrm UV+IR}$ relations, together with their 1-2$\sigma$ dispersion assuming $k_s=5$. The observed trends are comparable to those shown for $k_s=0.34$. The relations obtained for $k_s=5$ agree within the errors with the location onto the L$_{\text{CIII]}}-$SFR$_{\mathrm UV+IR}$ plane of $z\approx 3$ low-mass star-forming galaxies with stellar masses in the range $7.9<\log(M_{\star} /M_{\odot})<10.3$ selected from the CIII] $1909\mathring{A}$ emission by \citet{Llerena23}. This is true for all SFH scenarios examined in this work for BPASS, as well as in the case of a continuous SF assuming a SB99 spectrum.
In Figure \ref{fig:CIII]} we also include the $z\approx 6-7$ sources from \citet{Markov}, jointly detected in CIII] $1909\mathring{A}$ and [CII] 158$\rm \mu m$. \citet{Markov} data are systematically higher than our best-fit relations. This suggests that the three galaxies are experiencing starburst phase \citep[$k_s>5$, in line with the conclusions by ][]{Markov}.

At fixed SFR, the CIII] luminosity for $k_s=5$ is systematically higher than that obtained for $k_s=0.34$.
$k_s=0.34$ gives higher values of gas surface densities compared to $k_s=5$ (see Eq. (\ref{eq:ks})).
This results in a relatively high cloud density $n$ and a relatively lower ionisation parameter for $k_s=0.34$ compared to $k_s=5$ (see Eq. (\ref{eq:SAM})).
A small value of $U$ will produce a small HII region (see Figure \ref{fig:temp}), thus reducing the emission of UV and optical lines with high ionisation energies, as already remarked in Section \ref{sec:cloudy_output}.

In Figure \ref{fig:[OIII]} we show the theoretical predictions for the [OIII] $5007\rm \text{\AA}$ line for $k_s=0.34$ (left panel) and $k_s=5$ (right panel).  The data presented by \citet{Llerena23} are near the best-fit relation found with $k_s=5$, in the two SFH scenarios examined in the BPASS case, as well as in the case of a continuous SF for SB99. As for CIII], binary systems keep the brightness of the [OIII] line constant even if the galaxy is experiencing a quenching of $\Delta{t}_q=5$ Myr. SB99 quenched SF cases show instead a large drop in $L_{\rm [OIII]}$, as already underlined for CIII].
This is related to the ionisation potential of $O^+$ ($35.1\, \rm{eV}$) and to the shrinking of HII region due to a softer incident spectrum.

\begin{figure*}[htbp]
    \centering
    \includegraphics[width=8cm]{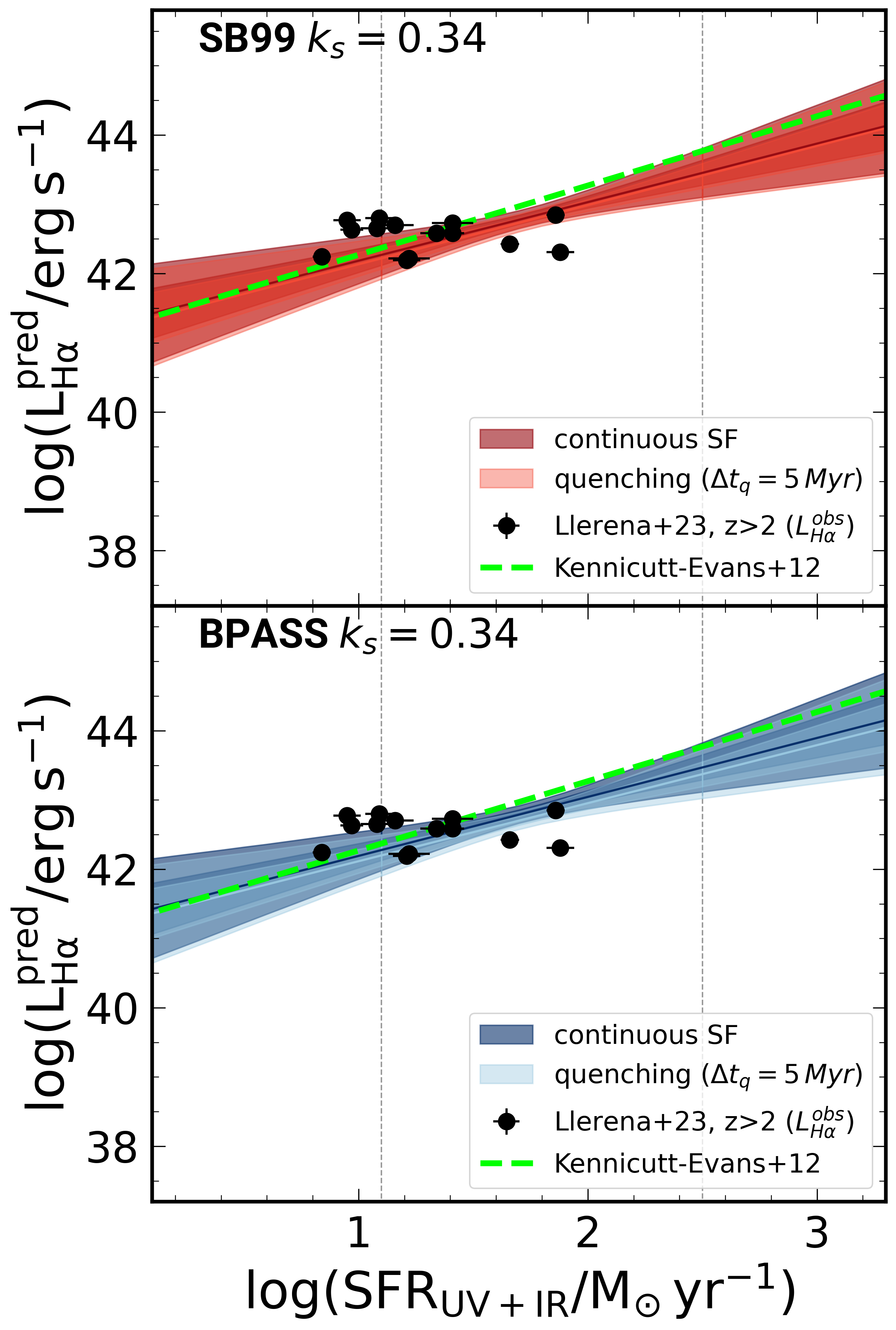}
    \includegraphics[width=8cm]{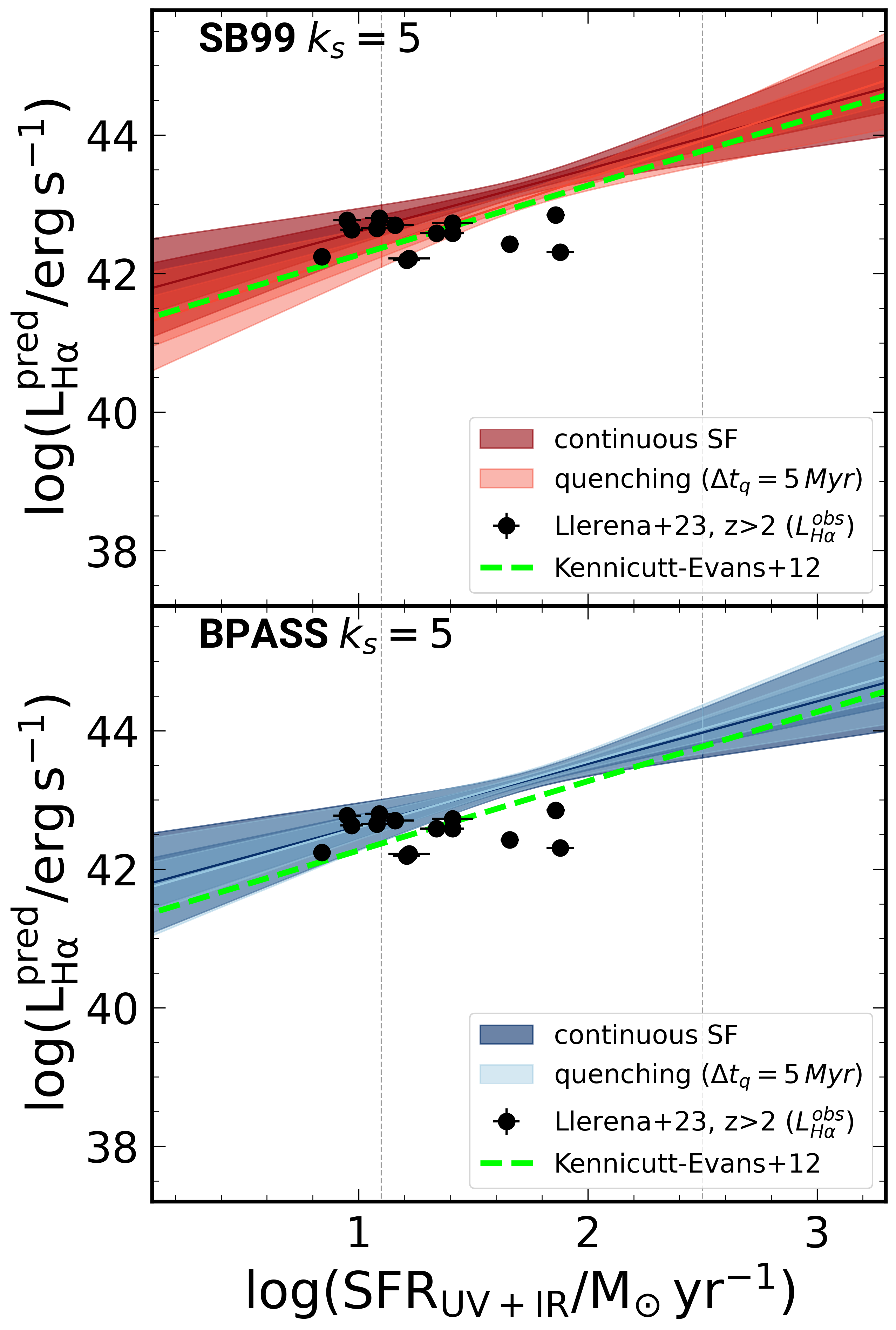}
    \caption{Predicted luminosity of H$\alpha$ $6563\rm \text{\AA}$ as a function of the total SFR (SFR$_{\rm UV+IR}$), for $k_s=0.34$ (left panel) and $k_s=5$ (right panel). The colours correspond to 1-2$\sigma$ dispersion. The central region between the grey dashed lines represents the SFR range of the ALPINE subsample modelled in this work. The red and blue colour palettes represent the two star formation histories considered in this work: continuous and $\Delta t_q=5\, \rm Myr$. 
    The black points are the observed data from\citet{Llerena23}.
    The green dashed line is the \citet{KE12} relation.}\label{fig:Ha}
\end{figure*}

In Figure \ref{fig:Ha} we show the predictions for the H$\alpha$ line. 
Contrary to CIII] and [OIII] $5007\rm \text{\AA}$ (Figures \ref{fig:CIII]}, \ref{fig:[OIII]}) the H$\alpha$ line does not show a marked dependence on the presence of a quenching in SF, at least up to $\Delta t_q=5\, \rm Myr$, in both the BPASS and SB99 cases. This is due to the behaviour of the cumulative flux of [CII] and H$\alpha$ in case of quenching discussed in Section \ref{sec:cloudy_output}.
For SB99, both lines show an enhancement in flux for the quenched case. The result is that the ratio between the two lines does not vary.
For BPASS, due to the presence of binaries (see Section \ref{sec:SB99vsBPASS}), the emission of the two lines is not affected by the quenching in SF, so the ratio remains constant.

H$\alpha$ predictions represent a key tool to verify whether our model can reproduce well-established relationships from the literature that link H$\alpha$ to the SFR of a galaxy \citep[e.g. ][]{Kennicutt,KE12}. As outlined in Figure \ref{fig:Ha} our model is able to reproduce these relationships, within a factor of 10.
For $k_s=0.34$ (left panel), we observe that the \citet{Kennicutt, KE12} relations agree within 1-2$\sigma$ dispersion of our modelling fits.
For $k_s=5$ (right panel) the predictions consistently exceed those from \citet{Kennicutt, KE12}, even remaining in the region of 2$\sigma$ dispersion.
These results suggest caution when applying locally calibrated relations to starburst galaxies.
The H$\alpha$ predictions are not significantly affected by an abrupt quenching  $\Delta t_q=5\, \rm Myr$ or by the presence of binary systems.
This aligns with the fact that 90\% of the H$\alpha$ emission originates from a stellar population with an average age of $10\,\rm Myr$, precluding observable luminosity variations in the analysed quenching scenario \citep[][]{KE12}.\\
Considering the BPASS models as fiducial given the inclusion of binary stars, the UV and optical lines are not affected by the SFH of a galaxy.
Instead, the line luminosity depends on the assumed value of the $k_s$ parameter (see Table \ref{table:RESULTS_1}).
From the position of an observed galaxy in the UV/optical-SFR relations we can therefore put a constraint about the galaxy's deviation from the local KS relation (i.e. an estimation for the $k_s$ parameter), as for instance the already proposed in \citet{Vallini+21, Vallini24} for FIR lines. For H$\beta$ and [NII] predictions, see Appendix \ref{sec:appendix}.

\section{Conclusions}\label{sec:conclusion}

In this work, we developed a physically motivated ISM model for the line emission from HII regions and PDR using \textsc{CLOUDY} \citep{Ferland} and two different codes for simulating the incident radiation from a stellar population: Starburst99 (SB99) and Binary Population and Spectral Synthesis Code (BPASS), with the latter including binary stars.  
By using the theoretical ratio from our model, we obtain predictions for the UV/optical line luminosity from the observed [CII] luminosity for a subsample of ALPINE galaxies. The main results can be summarised as follows:

\begin{itemize} 
    \item{The presence of binary stars as modelled by the BPASS code makes the UV spectrum from a composite stellar population harder with respect to what is obtained with SB99 without binary systems. This effect is particularly evident for a quenched SFH as binary stars delay the drop of the EUV photon flux after the quenching. The EUV flux is indeed sustained by UV photon production from mass-exchange processes in binary systems.
    These significantly affect the ionising photon rate and hence the size and temperature of the HII regions. Assuming an abrupt SF quenching for SB99 models, the decrease of EUV photon budget in the incident radiation reduces the width and alters the chemical properties of the HII region.
    Instead, binary stars included in the BPASS code are able to sustain emission in EUV after quenching, thus keeping the HII region unchanged for at least $\Delta t_q=5\, \rm Myr$. Therefore, the presence of binaries must be considered in a complete line emission modelling (as e.g. suggested by \citealt{Jaskot16}).} 
    \item{We derive the ISM physical properties (hydrogen density $n$ and ionisation parameter $U$) of a subsample of 44 ALPINE galaxies.
    We use the analytical model from \citet{Ferrara} to associate $n$ and~$U$, to the gas surface density ($\Sigma_{\rm gas}$), obtained from the KS relation, and the star formation surface density ($\Sigma_{\rm SFR}$) of each galaxy, through the burstiness parameter $k_s$.
    For the ALPINE subsample we consider a fiducial value of $k_s=0.34$, estimated from \citet{Bethermin24} data.
    We obtain $\log U_{0.34}$ ($-3.8\pm 0.2$) and $\log n_{0.34}$ ($2.9\pm 0.6$).
    These values result to be consistent with those inferred by\citet{Vanderhoof+22}, who investigated a smaller subset of ALPINE galaxies.}
    \item{We derive UV/optical lines luminosities$-$SFR relations, in the form $    \log(L_{line}/\rm erg\, s^{-1})=\alpha\log(\rm SFR/M_{\odot}\, yr^{-1})+\beta,$ for different $k_s$ values: the fiducial for ALPINE galaxies ($k_s\approx 0.34$) and $k_s=1,5,10$ (see Table \ref{table:RESULTS_1}). The relations, in line with the high-$z$ data from \citet{Llerena23}, suggest their potential use at high-$z$ for galaxy SFR characterisation. The strong alignment of the \citet{Kennicutt,KE12} relations with our model predictions, for $k_s=0.34$, attests to our models' robustness. 
    For starburst galaxies ($k_s>1$), our predictions exceed those of the Kennicutt relation, suggesting caution when applying locally calibrated Kennicutt laws to highly efficient star-forming galaxies. }
    \item{In the models that use the spectral synthesis code BPASS for estimating the radiation from the stellar population, the relations have negligible SFH dependence but show a strong dependence on the $k_s$ value, with line luminosities increasing for higher $k_s$, namely for starburst galaxies.
    For the SB99 cases, the key dependence is on SFH, and the lack of binaries results in a significant flux drop even after a short quenching period ($\Delta t_q=5\, \rm Myr$) for at least UV an optical line with high ionisation energy.}  
\end{itemize}

Our method represents an important tool to provide stringent constraints on the UV/optical lines luminosities, drawing from the robust [CII] ALMA observations. Looking ahead, the future JWST follow-up of ALMA observations promises to provide state of art data to these predictions, testing their efficacy in characterising the SFR of high-$z$ galaxies, and opening up an opportunity to characterise the star formation mode in galaxies at the end of the EoR.
In a forthcoming paper the imminent JWST NIRSpec IFU observations of 18 ALPINE galaxies will offer a great opportunity to exploit our theoretical predictions thereby potentially confirming or not the physical properties found for these galaxies.

\begin{acknowledgements}
We thank the anonymous referee for their constructive report, which helped to improve the quality of this work. 
MB gratefully acknowledges support from the ANID BASAL project FB210003 and from the FONDECYT regular grant 1211000.
This work was supported by the French government through the France 2030 investment plan managed by the National Research Agency (ANR), as part of the Initiative of Excellence of Université Côte d’Azur under reference number ANR-15-IDEX-01.
MR acknowledges support from the Narodowe Centrum Nauki (UMO-2020/38/E/ST9/00077) and support from the Foundation for Polish Science (FNP) under the program START 063.2023.
EI acknowledge funding by ANID FONDECYT Regular 1221846.
FE acknowledges support from grant PRIN MIUR 2017-20173ML3WW\_001. LV, FE acknowledge funding from the INAF Mini Grant 2022 program “Face-to-Face with the Local Universe: ISM Empowerment (LOCAL)”.
\end{acknowledgements}

\bibliographystyle{mnras}
\bibliography{bibliography}

\begin{thebibliography}{}
\makeatletter
\relax
\def\mn@urlcharsother{\let\do\@makeother \do\$\do\&\do\#\do\^\do\_\do\%\do\~}
\def\mn@doi{\begingroup\mn@urlcharsother \@ifnextchar [ {\mn@doi@} {\mn@doi@[]}}
\def\mn@doi@[#1]#2{\def\@tempa{#1}\ifx\@tempa\@empty \href {http://dx.doi.org/#2} {doi:#2}\else \href {http://dx.doi.org/#2} {#1}\fi \endgroup}
\def\mn@eprint#1#2{\mn@eprint@#1:#2::\@nil}
\def\mn@eprint@arXiv#1{\href {http://arxiv.org/abs/#1} {{\tt arXiv:#1}}}
\def\mn@eprint@dblp#1{\href {http://dblp.uni-trier.de/rec/bibtex/#1.xml} {dblp:#1}}
\def\mn@eprint@#1:#2:#3:#4\@nil{\def\@tempa {#1}\def\@tempb {#2}\def\@tempc {#3}\ifx \@tempc \@empty \let \@tempc \@tempb \let \@tempb \@tempa \fi \ifx \@tempb \@empty \def\@tempb {arXiv}\fi \@ifundefined {mn@eprint@\@tempb}{\@tempb:\@tempc}{\expandafter \expandafter \csname mn@eprint@\@tempb\endcsname \expandafter{\@tempc}}}

\bibitem[\protect\citeauthoryear{{Arata}, {Yajima}, {Nagamine}, {Abe}  \& {Khochfar}}{{Arata} et~al.}{2020}]{Arata+20}
{Arata} S.,  {Yajima} H.,  {Nagamine} K.,  {Abe} M.,   {Khochfar} S.,  2020, \mn@doi [\mnras] {10.1093/mnras/staa2809}, \href {https://ui.adsabs.harvard.edu/abs/2020MNRAS.498.5541A} {498, 5541}

\bibitem[\protect\citeauthoryear{{Backhaus} et~al.,}{{Backhaus} et~al.}{2024}]{Backhaus+24}
{Backhaus} B.~E.,  et~al., 2024, \mn@doi [\apj] {10.3847/1538-4357/ad1520}, \href {https://ui.adsabs.harvard.edu/abs/2024ApJ...962..195B} {962, 195}

\bibitem[\protect\citeauthoryear{{Baldwin}, {Ferland}, {Martin}, {Corbin}, {Cota}, {Peterson}  \& {Slettebak}}{{Baldwin} et~al.}{1991}]{Baldwin+91}
{Baldwin} J.~A.,  {Ferland} G.~J.,  {Martin} P.~G.,  {Corbin} M.~R.,  {Cota} S.~A.,  {Peterson} B.~M.,   {Slettebak} A.,  1991, \mn@doi [\apj] {10.1086/170146}, \href {https://ui.adsabs.harvard.edu/abs/1991ApJ...374..580B} {374, 580}

\bibitem[\protect\citeauthoryear{{Barchiesi} et~al.,}{{Barchiesi} et~al.}{2023}]{Barchiesi+23}
{Barchiesi} L.,  et~al., 2023, \mn@doi [\aap] {10.1051/0004-6361/202244838}, \href {https://ui.adsabs.harvard.edu/abs/2023A&A...675A..30B} {675, A30}

\bibitem[\protect\citeauthoryear{{Behrens}, {Pallottini}, {Ferrara}, {Gallerani}  \& {Vallini}}{{Behrens} et~al.}{2018}]{Behrens+18}
{Behrens} C.,  {Pallottini} A.,  {Ferrara} A.,  {Gallerani} S.,   {Vallini} L.,  2018, \mn@doi [\mnras] {10.1093/mnras/sty552}, \href {https://ui.adsabs.harvard.edu/abs/2018MNRAS.477..552B} {477, 552}

\bibitem[\protect\citeauthoryear{{B{\'e}thermin} et~al.,}{{B{\'e}thermin} et~al.}{2020a}]{Bethermin+20}
{B{\'e}thermin} M.,  et~al., 2020a, \mn@doi [\aap] {10.1051/0004-6361/202037649}, \href {https://ui.adsabs.harvard.edu/abs/2020A&A...643A...2B} {643, A2}

\bibitem[\protect\citeauthoryear{{B{\'e}thermin} et~al.,}{{B{\'e}thermin} et~al.}{2020b}]{Bethermin}
{B{\'e}thermin} M.,  et~al., 2020b, \mn@doi [\aap] {10.1051/0004-6361/202037649}, \href {https://ui.adsabs.harvard.edu/abs/2020A&A...643A...2B} {643, A2}

\bibitem[\protect\citeauthoryear{{B{\'e}thermin} et~al.,}{{B{\'e}thermin} et~al.}{2023}]{Bethermin24}
{B{\'e}thermin} M.,  et~al., 2023, \mn@doi [\aap] {10.1051/0004-6361/202348115}, \href {https://ui.adsabs.harvard.edu/abs/2023A&A...680L...8B} {680, L8}

\bibitem[\protect\citeauthoryear{{Bouwens} et~al.,}{{Bouwens} et~al.}{2022}]{Bouwens}
{Bouwens} R.~J.,  et~al., 2022, \mn@doi [\apj] {10.3847/1538-4357/ac5a4a}, \href {https://ui.adsabs.harvard.edu/abs/2022ApJ...931..160B} {931, 160}

\bibitem[\protect\citeauthoryear{Bunker et~al.,}{Bunker et~al.}{2023}]{Bunker+23}
Bunker A.~J.,  et~al., 2023, \mn@doi [Astronomy \&amp; Astrophysics] {10.1051/0004-6361/202346159}, 677, A88

\bibitem[\protect\citeauthoryear{{Carilli} \& {Walter}}{{Carilli} \& {Walter}}{2013}]{Carilli2013}
{Carilli} C.~L.,  {Walter} F.,  2013, \mn@doi [\araa] {10.1146/annurev-astro-082812-140953}, \href {https://ui.adsabs.harvard.edu/abs/2013ARA&A..51..105C} {51, 105}

\bibitem[\protect\citeauthoryear{{Carniani}, {Maiolino}, {Smit}  \& {Amor{\'\i}n}}{{Carniani} et~al.}{2018a}]{Carniani}
{Carniani} S.,  {Maiolino} R.,  {Smit} R.,   {Amor{\'\i}n} R.,  2018a, \mn@doi [\apjl] {10.3847/2041-8213/aaab45}, \href {https://ui.adsabs.harvard.edu/abs/2018ApJ...854L...7C} {854, L7}

\bibitem[\protect\citeauthoryear{{Carniani}, {Maiolino}, {Smit}  \& {Amor{\'\i}n}}{{Carniani} et~al.}{2018b}]{Carniani+18}
{Carniani} S.,  {Maiolino} R.,  {Smit} R.,   {Amor{\'\i}n} R.,  2018b, \mn@doi [\apjl] {10.3847/2041-8213/aaab45}, \href {https://ui.adsabs.harvard.edu/abs/2018ApJ...854L...7C} {854, L7}

\bibitem[\protect\citeauthoryear{{Carniani} et~al.,}{{Carniani} et~al.}{2024}]{Carniani+24}
{Carniani} S.,  et~al., 2024, \mn@doi [\nat] {10.1038/s41586-024-07860-9}, \href {https://ui.adsabs.harvard.edu/abs/2024Natur.633..318C} {633, 318}

\bibitem[\protect\citeauthoryear{{Chabrier}}{{Chabrier}}{2003}]{Chabrier+03}
{Chabrier} G.,  2003, \mn@doi [\pasp] {10.1086/376392}, \href {https://ui.adsabs.harvard.edu/abs/2003PASP..115..763C} {115, 763}

\bibitem[\protect\citeauthoryear{{Chatzikos} et~al.,}{{Chatzikos} et~al.}{2023}]{Cloudy+23}
{Chatzikos} M.,  et~al., 2023, \mn@doi [\rmxaa] {10.22201/ia.01851101p.2023.59.02.12}, \href {https://ui.adsabs.harvard.edu/abs/2023RMxAA..59..327C} {59, 327}

\bibitem[\protect\citeauthoryear{{Curti} et~al.,}{{Curti} et~al.}{2023}]{curti2023}
{Curti} M.,  et~al., 2023, \mn@doi [\mnras] {10.1093/mnras/stac2737}, \href {https://ui.adsabs.harvard.edu/abs/2023MNRAS.518..425C} {518, 425}

\bibitem[\protect\citeauthoryear{{Dayal} \& {Ferrara}}{{Dayal} \& {Ferrara}}{2018}]{Dayal+18}
{Dayal} P.,  {Ferrara} A.,  2018, \mn@doi [\physrep] {10.1016/j.physrep.2018.10.002}, \href {https://ui.adsabs.harvard.edu/abs/2018PhR...780....1D} {780, 1}

\bibitem[\protect\citeauthoryear{{De Looze} et~al.,}{{De Looze} et~al.}{2014}]{DeLooze}
{De Looze} I.,  et~al., 2014, \mn@doi [\aap] {10.1051/0004-6361/201322489}, \href {https://ui.adsabs.harvard.edu/abs/2014A&A...568A..62D} {568, A62}

\bibitem[\protect\citeauthoryear{{Dessauges-Zavadsky} et~al.,}{{Dessauges-Zavadsky} et~al.}{2020}]{DZ}
{Dessauges-Zavadsky} M.,  et~al., 2020, \mn@doi [\aap] {10.1051/0004-6361/202038231}, \href {https://ui.adsabs.harvard.edu/abs/2020A&A...643A...5D} {643, A5}

\bibitem[\protect\citeauthoryear{{Di Cesare}, {Graziani}, {Schneider}, {Ginolfi}, {Venditti}, {Santini}  \& {Hunt}}{{Di Cesare} et~al.}{2023}]{DiCesare+23}
{Di Cesare} C.,  {Graziani} L.,  {Schneider} R.,  {Ginolfi} M.,  {Venditti} A.,  {Santini} P.,   {Hunt} L.~K.,  2023, \mn@doi [\mnras] {10.1093/mnras/stac3702}, \href {https://ui.adsabs.harvard.edu/abs/2023MNRAS.519.4632D} {519, 4632}

\bibitem[\protect\citeauthoryear{{Donnan} et~al.,}{{Donnan} et~al.}{2024}]{Donnan+24}
{Donnan} C.~T.,  et~al., 2024, \mn@doi [\mnras] {10.1093/mnras/stae2037}, \href {https://ui.adsabs.harvard.edu/abs/2024MNRAS.533.3222D} {533, 3222}

\bibitem[\protect\citeauthoryear{{Dunlop} et~al.,}{{Dunlop} et~al.}{2021}]{Dunlop+21}
{Dunlop} J.~S.,  et~al., 2021, {PRIMER: Public Release IMaging for Extragalactic Research}, JWST Proposal. Cycle 1, ID. \#1837

\bibitem[\protect\citeauthoryear{{Eisenstein} et~al.,}{{Eisenstein} et~al.}{2023}]{Eisenstein}
{Eisenstein} D.~J.,  et~al., 2023, \mn@doi [\apjs] {10.48550/arXiv.2306.02465}, \href {https://ui.adsabs.harvard.edu/abs/2023arXiv230602465E} {p. to appear}

\bibitem[\protect\citeauthoryear{{Eldridge} \& {Stanway}}{{Eldridge} \& {Stanway}}{2022}]{Eldridge2022}
{Eldridge} J.~J.,  {Stanway} E.~R.,  2022, \mn@doi [\araa] {10.1146/annurev-astro-052920-100646}, \href {https://ui.adsabs.harvard.edu/abs/2022ARA&A..60..455E} {60, 455}

\bibitem[\protect\citeauthoryear{{Faisst} et~al.,}{{Faisst} et~al.}{2016}]{Faisst+16}
{Faisst} A.~L.,  et~al., 2016, \mn@doi [\apj] {10.3847/0004-637X/822/1/29}, \href {https://ui.adsabs.harvard.edu/abs/2016ApJ...822...29F} {822, 29}

\bibitem[\protect\citeauthoryear{{Faisst} et~al.,}{{Faisst} et~al.}{2020}]{Faisst}
{Faisst} A.~L.,  et~al., 2020, \mn@doi [\apjs] {10.3847/1538-4365/ab7ccd}, \href {https://ui.adsabs.harvard.edu/abs/2020ApJS..247...61F} {247, 61}

\bibitem[\protect\citeauthoryear{{Feltre}, {Charlot}  \& {Gutkin}}{{Feltre} et~al.}{2016}]{Feltre+16}
{Feltre} A.,  {Charlot} S.,   {Gutkin} J.,  2016, \mn@doi [\mnras] {10.1093/mnras/stv2794}, \href {https://ui.adsabs.harvard.edu/abs/2016MNRAS.456.3354F} {456, 3354}

\bibitem[\protect\citeauthoryear{{Ferland} et~al.,}{{Ferland} et~al.}{2017}]{Ferland}
{Ferland} G.~J.,  et~al., 2017, \mn@doi [\rmxaa] {10.48550/arXiv.1705.10877}, \href {https://ui.adsabs.harvard.edu/abs/2017RMxAA..53..385F} {53, 385}

\bibitem[\protect\citeauthoryear{{Ferrara}, {Vallini}, {Pallottini}, {Gallerani}, {Carniani}, {Kohandel}, {Decataldo}  \& {Behrens}}{{Ferrara} et~al.}{2019}]{Ferrara}
{Ferrara} A.,  {Vallini} L.,  {Pallottini} A.,  {Gallerani} S.,  {Carniani} S.,  {Kohandel} M.,  {Decataldo} D.,   {Behrens} C.,  2019, \mn@doi [\mnras] {10.1093/mnras/stz2031}, \href {https://ui.adsabs.harvard.edu/abs/2019MNRAS.489....1F} {489, 1}

\bibitem[\protect\citeauthoryear{{Foreman-Mackey}, {Hogg}, {Lang}  \& {Goodman}}{{Foreman-Mackey} et~al.}{2013}]{emcee}
{Foreman-Mackey} D.,  {Hogg} D.~W.,  {Lang} D.,   {Goodman} J.,  2013, \mn@doi [\pasp] {10.1086/670067}, \href {https://ui.adsabs.harvard.edu/abs/2013PASP..125..306F} {125, 306}

\bibitem[\protect\citeauthoryear{{Fudamoto} et~al.,}{{Fudamoto} et~al.}{2020}]{Fudamoto+20}
{Fudamoto} Y.,  et~al., 2020, \mn@doi [\aap] {10.1051/0004-6361/202038163}, \href {https://ui.adsabs.harvard.edu/abs/2020A&A...643A...4F} {643, A4}

\bibitem[\protect\citeauthoryear{{Fujimoto} et~al.,}{{Fujimoto} et~al.}{2020}]{Fujimoto}
{Fujimoto} S.,  et~al., 2020, \mn@doi [\apj] {10.3847/1538-4357/ab94b3}, \href {https://ui.adsabs.harvard.edu/abs/2020ApJ...900....1F} {900, 1}

\bibitem[\protect\citeauthoryear{{Giacconi} et~al.,}{{Giacconi} et~al.}{2002}]{Giacconi+02}
{Giacconi} R.,  et~al., 2002, \mn@doi [\apjs] {10.1086/338927}, \href {https://ui.adsabs.harvard.edu/abs/2002ApJS..139..369G} {139, 369}

\bibitem[\protect\citeauthoryear{{Grevesse}, {Asplund}, {Sauval}  \& {Scott}}{{Grevesse} et~al.}{2010}]{Grevesse}
{Grevesse} N.,  {Asplund} M.,  {Sauval} A.~J.,   {Scott} P.,  2010, \mn@doi [\apss] {10.1007/s10509-010-0288-z}, \href {https://ui.adsabs.harvard.edu/abs/2010Ap&SS.328..179G} {328, 179}

\bibitem[\protect\citeauthoryear{Götberg, de Mink, Groh, Leitherer  \& Norman}{Götberg et~al.}{2019}]{Gotberg+19}
Götberg Y.,  de Mink S.~E.,  Groh J.~H.,  Leitherer C.,   Norman C.,  2019, \mn@doi [Astronomy \&amp; Astrophysics] {10.1051/0004-6361/201834525}, 629, A134

\bibitem[\protect\citeauthoryear{{Heiderman}, {Evans}, {Allen}, {Huard}  \& {Heyer}}{{Heiderman} et~al.}{2010}]{Heiderman+10}
{Heiderman} A.,  {Evans} Neal~J. I.,  {Allen} L.~E.,  {Huard} T.,   {Heyer} M.,  2010, \mn@doi [\apj] {10.1088/0004-637X/723/2/1019}, \href {https://ui.adsabs.harvard.edu/abs/2010ApJ...723.1019H} {723, 1019}

\bibitem[\protect\citeauthoryear{{Indriolo}, {Geballe}, {Oka}  \& {McCall}}{{Indriolo} et~al.}{2007}]{Indriolo}
{Indriolo} N.,  {Geballe} T.~R.,  {Oka} T.,   {McCall} B.~J.,  2007, \mn@doi [\apj] {10.1086/523036}, \href {https://ui.adsabs.harvard.edu/abs/2007ApJ...671.1736I} {671, 1736}

\bibitem[\protect\citeauthoryear{{Isobe}, {Ouchi}, {Nakajima}, {Harikane}, {Ono}, {Xu}, {Zhang}  \& {Umeda}}{{Isobe} et~al.}{2023}]{isobe2023}
{Isobe} Y.,  {Ouchi} M.,  {Nakajima} K.,  {Harikane} Y.,  {Ono} Y.,  {Xu} Y.,  {Zhang} Y.,   {Umeda} H.,  2023, \mn@doi [\apj] {10.3847/1538-4357/acf376}, \href {https://ui.adsabs.harvard.edu/abs/2023ApJ...956..139I} {956, 139}

\bibitem[\protect\citeauthoryear{{Jaskot} \& {Ravindranath}}{{Jaskot} \& {Ravindranath}}{2016a}]{Jaskot16}
{Jaskot} A.~E.,  {Ravindranath} S.,  2016a, \mn@doi [\apj] {10.3847/1538-4357/833/2/136}, \href {https://ui.adsabs.harvard.edu/abs/2016ApJ...833..136J} {833, 136}

\bibitem[\protect\citeauthoryear{{Jaskot} \& {Ravindranath}}{{Jaskot} \& {Ravindranath}}{2016b}]{Jaskot+16}
{Jaskot} A.~E.,  {Ravindranath} S.,  2016b, \mn@doi [\apj] {10.3847/1538-4357/833/2/136}, \href {https://ui.adsabs.harvard.edu/abs/2016ApJ...833..136J} {833, 136}

\bibitem[\protect\citeauthoryear{{Jones} et~al.,}{{Jones} et~al.}{2021}]{Jones+21}
{Jones} G.~C.,  et~al., 2021, \mn@doi [\mnras] {10.1093/mnras/stab2226}, \href {https://ui.adsabs.harvard.edu/abs/2021MNRAS.507.3540J} {507, 3540}

\bibitem[\protect\citeauthoryear{{Karttunen}}{{Karttunen}}{2007}]{Kartunnen}
{Karttunen} 2007, {Fundamental Astronomy}.
Springer

\bibitem[\protect\citeauthoryear{{Katz}, {Kimm}, {Sijacki}  \& {Haehnelt}}{{Katz} et~al.}{2017}]{Katz+17}
{Katz} H.,  {Kimm} T.,  {Sijacki} D.,   {Haehnelt} M.~G.,  2017, \mn@doi [\mnras] {10.1093/mnras/stx608}, \href {https://ui.adsabs.harvard.edu/abs/2017MNRAS.468.4831K} {468, 4831}

\bibitem[\protect\citeauthoryear{{Katz} et~al.,}{{Katz} et~al.}{2022}]{Katz+22}
{Katz} H.,  et~al., 2022, \mn@doi [\mnras] {10.1093/mnras/stac028}, \href {https://ui.adsabs.harvard.edu/abs/2022MNRAS.510.5603K} {510, 5603}

\bibitem[\protect\citeauthoryear{{Kennicutt}}{{Kennicutt}}{1998}]{Kennicutt}
{Kennicutt} Robert~C. J.,  1998, \mn@doi [\araa] {10.1146/annurev.astro.36.1.189}, \href {https://ui.adsabs.harvard.edu/abs/1998ARA&A..36..189K} {36, 189}

\bibitem[\protect\citeauthoryear{{Kennicutt} \& {Evans}}{{Kennicutt} \& {Evans}}{2012}]{KE12}
{Kennicutt} R.~C.,  {Evans} N.~J.,  2012, \mn@doi [\araa] {10.1146/annurev-astro-081811-125610}, \href {https://ui.adsabs.harvard.edu/abs/2012ARA&A..50..531K} {50, 531}

\bibitem[\protect\citeauthoryear{{Kewley}, {Nicholls}  \& {Sutherland}}{{Kewley} et~al.}{2019}]{kewley2019}
{Kewley} L.~J.,  {Nicholls} D.~C.,   {Sutherland} R.~S.,  2019, \mn@doi [\araa] {10.1146/annurev-astro-081817-051832}, \href {https://ui.adsabs.harvard.edu/abs/2019ARA&A..57..511K} {57, 511}

\bibitem[\protect\citeauthoryear{{Kohandel}, {Pallottini}, {Ferrara}, {Carniani}, {Gallerani}, {Vallini}, {Zanella}  \& {Behrens}}{{Kohandel} et~al.}{2020}]{Kohandel+20}
{Kohandel} M.,  {Pallottini} A.,  {Ferrara} A.,  {Carniani} S.,  {Gallerani} S.,  {Vallini} L.,  {Zanella} A.,   {Behrens} C.,  2020, \mn@doi [\mnras] {10.1093/mnras/staa2792}, \href {https://ui.adsabs.harvard.edu/abs/2020MNRAS.499.1250K} {499, 1250}

\bibitem[\protect\citeauthoryear{{Kriek} et~al.,}{{Kriek} et~al.}{2015}]{Kriek+15}
{Kriek} M.,  et~al., 2015, \mn@doi [\apjs] {10.1088/0067-0049/218/2/15}, \href {https://ui.adsabs.harvard.edu/abs/2015ApJS..218...15K} {218, 15}

\bibitem[\protect\citeauthoryear{{Lagache}, {Cousin}  \& {Chatzikos}}{{Lagache} et~al.}{2018}]{Lagache+18}
{Lagache} G.,  {Cousin} M.,   {Chatzikos} M.,  2018, \mn@doi [\aap] {10.1051/0004-6361/201732019}, \href {https://ui.adsabs.harvard.edu/abs/2018A&A...609A.130L} {609, A130}

\bibitem[\protect\citeauthoryear{{Le F{\`e}vre} et~al.,}{{Le F{\`e}vre} et~al.}{2020}]{Fevre}
{Le F{\`e}vre} O.,  et~al., 2020, \mn@doi [\aap] {10.1051/0004-6361/201936965}, \href {https://ui.adsabs.harvard.edu/abs/2020A&A...643A...1L} {643, A1}

\bibitem[\protect\citeauthoryear{{Leitherer} et~al.,}{{Leitherer} et~al.}{1999}]{SB99}
{Leitherer} C.,  et~al., 1999, \mn@doi [\apjs] {10.1086/313233}, \href {https://ui.adsabs.harvard.edu/abs/1999ApJS..123....3L} {123, 3}

\bibitem[\protect\citeauthoryear{{Lelli}, {Di Teodoro}, {Fraternali}, {Man}, {Zhang}, {De Breuck}, {Davis}  \& {Maiolino}}{{Lelli} et~al.}{2021}]{Lelli+21}
{Lelli} F.,  {Di Teodoro} E.~M.,  {Fraternali} F.,  {Man} A. W.~S.,  {Zhang} Z.-Y.,  {De Breuck} C.,  {Davis} T.~A.,   {Maiolino} R.,  2021, \mn@doi [Science] {10.1126/science.abc1893}, \href {https://ui.adsabs.harvard.edu/abs/2021Sci...371..713L} {371, 713}

\bibitem[\protect\citeauthoryear{{Llerena} et~al.,}{{Llerena} et~al.}{2023}]{Llerena23}
{Llerena} M.,  et~al., 2023, \mn@doi [\aap] {10.1051/0004-6361/202346232}, \href {https://ui.adsabs.harvard.edu/abs/2023A&A...676A..53L} {676, A53}

\bibitem[\protect\citeauthoryear{{Lupi}, {Pallottini}, {Ferrara}, {Bovino}, {Carniani}  \& {Vallini}}{{Lupi} et~al.}{2020}]{Lupi+20}
{Lupi} A.,  {Pallottini} A.,  {Ferrara} A.,  {Bovino} S.,  {Carniani} S.,   {Vallini} L.,  2020, \mn@doi [\mnras] {10.1093/mnras/staa1842}, \href {https://ui.adsabs.harvard.edu/abs/2020MNRAS.496.5160L} {496, 5160}

\bibitem[\protect\citeauthoryear{{Maiolino} et~al.,}{{Maiolino} et~al.}{2015}]{Maiolino}
{Maiolino} R.,  et~al., 2015, \mn@doi [\mnras] {10.1093/mnras/stv1194}, \href {https://ui.adsabs.harvard.edu/abs/2015MNRAS.452...54M} {452, 54}

\bibitem[\protect\citeauthoryear{{Markov}, {Carniani}, {Vallini}, {Ferrara}, {Pallottini}, {Maiolino}, {Gallerani}  \& {Pentericci}}{{Markov} et~al.}{2022}]{Markov}
{Markov} V.,  {Carniani} S.,  {Vallini} L.,  {Ferrara} A.,  {Pallottini} A.,  {Maiolino} R.,  {Gallerani} S.,   {Pentericci} L.,  2022, \mn@doi [\aap] {10.1051/0004-6361/202243336}, \href {https://ui.adsabs.harvard.edu/abs/2022A&A...663A.172M} {663, A172}

\bibitem[\protect\citeauthoryear{{Mather}, {Fixsen}, {Shafer}, {Mosier}  \& {Wilkinson}}{{Mather} et~al.}{1999}]{Mather+99}
{Mather} J.~C.,  {Fixsen} D.~J.,  {Shafer} R.~A.,  {Mosier} C.,   {Wilkinson} D.~T.,  1999, \mn@doi [\apj] {10.1086/306805}, \href {https://ui.adsabs.harvard.edu/abs/1999ApJ...512..511M} {512, 511}

\bibitem[\protect\citeauthoryear{{Mathis}, {Rumpl}  \& {Nordsieck}}{{Mathis} et~al.}{1977}]{Mathis+77}
{Mathis} J.~S.,  {Rumpl} W.,   {Nordsieck} K.~H.,  1977, \mn@doi [\apj] {10.1086/155591}, \href {https://ui.adsabs.harvard.edu/abs/1977ApJ...217..425M} {217, 425}

\bibitem[\protect\citeauthoryear{{Matthee} et~al.,}{{Matthee} et~al.}{2019}]{Matthee+19}
{Matthee} J.,  et~al., 2019, \mn@doi [\apj] {10.3847/1538-4357/ab2f81}, \href {https://ui.adsabs.harvard.edu/abs/2019ApJ...881..124M} {881, 124}

\bibitem[\protect\citeauthoryear{{McLure} et~al.,}{{McLure} et~al.}{2018}]{McLure+18}
{McLure} R.~J.,  et~al., 2018, \mn@doi [\mnras] {10.1093/mnras/sty1213}, \href {https://ui.adsabs.harvard.edu/abs/2018MNRAS.479...25M} {479, 25}

\bibitem[\protect\citeauthoryear{{Mitsuhashi} et~al.,}{{Mitsuhashi} et~al.}{2024}]{Mitsuhashi}
{Mitsuhashi} I.,  et~al., 2024, \mn@doi [\aap] {10.1051/0004-6361/202348782}, \href {https://ui.adsabs.harvard.edu/abs/2024A&A...690A.197M} {690, A197}

\bibitem[\protect\citeauthoryear{{Moriwaki} et~al.,}{{Moriwaki} et~al.}{2018}]{Moriwaki+18}
{Moriwaki} K.,  et~al., 2018, \mn@doi [\mnras] {10.1093/mnrasl/sly167}, \href {https://ui.adsabs.harvard.edu/abs/2018MNRAS.481L..84M} {481, L84}

\bibitem[\protect\citeauthoryear{{Nakajima} et~al.,}{{Nakajima} et~al.}{2018}]{Nakajima+18}
{Nakajima} K.,  et~al., 2018, \mn@doi [\aap] {10.1051/0004-6361/201731935}, \href {https://ui.adsabs.harvard.edu/abs/2018A&A...612A..94N} {612, A94}

\bibitem[\protect\citeauthoryear{{Nakazato}, {Yoshida}  \& {Ceverino}}{{Nakazato} et~al.}{2023}]{Nakazato+23}
{Nakazato} Y.,  {Yoshida} N.,   {Ceverino} D.,  2023, \mn@doi [\apj] {10.3847/1538-4357/ace25a}, \href {https://ui.adsabs.harvard.edu/abs/2023ApJ...953..140N} {953, 140}

\bibitem[\protect\citeauthoryear{{Osterbrock} \& {Ferland}}{{Osterbrock} \& {Ferland}}{2006}]{Ostembrook2006}
{Osterbrock} D.~E.,  {Ferland} G.~J.,  2006, {Astrophysics of gaseous nebulae and active galactic nuclei}.
University Science Books

\bibitem[\protect\citeauthoryear{{Pallottini} et~al.,}{{Pallottini} et~al.}{2019}]{Pallottini+19}
{Pallottini} A.,  et~al., 2019, \mn@doi [\mnras] {10.1093/mnras/stz1383}, \href {https://ui.adsabs.harvard.edu/abs/2019MNRAS.487.1689P} {487, 1689}

\bibitem[\protect\citeauthoryear{{Pallottini} et~al.,}{{Pallottini} et~al.}{2022}]{Pallottini+22}
{Pallottini} A.,  et~al., 2022, \mn@doi [\mnras] {10.1093/mnras/stac1281}, \href {https://ui.adsabs.harvard.edu/abs/2022MNRAS.513.5621P} {513, 5621}

\bibitem[\protect\citeauthoryear{{Parlanti}, {Carniani}, {Pallottini}, {Cignoni}, {Cresci}, {Kohandel}, {Mannucci}  \& {Marconi}}{{Parlanti} et~al.}{2023}]{Parlanti+23}
{Parlanti} E.,  {Carniani} S.,  {Pallottini} A.,  {Cignoni} M.,  {Cresci} G.,  {Kohandel} M.,  {Mannucci} F.,   {Marconi} A.,  2023, \mn@doi [\aap] {10.1051/0004-6361/202245603}, \href {https://ui.adsabs.harvard.edu/abs/2023A&A...673A.153P} {673, A153}

\bibitem[\protect\citeauthoryear{{Pozzi} et~al.,}{{Pozzi} et~al.}{2021}]{Pozzi}
{Pozzi} F.,  et~al., 2021, \mn@doi [\aap] {10.1051/0004-6361/202040258}, \href {https://ui.adsabs.harvard.edu/abs/2021A&A...653A..84P} {653, A84}

\bibitem[\protect\citeauthoryear{{Pozzi} et~al.,}{{Pozzi} et~al.}{2024}]{Pozzi+24}
{Pozzi} F.,  et~al., 2024, \mn@doi [\aap] {10.1051/0004-6361/202348996}, \href {https://ui.adsabs.harvard.edu/abs/2024A&A...686A.187P} {686, A187}

\bibitem[\protect\citeauthoryear{{Rieke} et~al.,}{{Rieke} et~al.}{2023}]{Rieke+23}
{Rieke} M.~J.,  et~al., 2023, \mn@doi [\apjs] {10.3847/1538-4365/acf44d}, \href {https://ui.adsabs.harvard.edu/abs/2023ApJS..269...16R} {269, 16}

\bibitem[\protect\citeauthoryear{{Rizzo}, {Vegetti}, {Fraternali}, {Stacey}  \& {Powell}}{{Rizzo} et~al.}{2021}]{Rizzo+21}
{Rizzo} F.,  {Vegetti} S.,  {Fraternali} F.,  {Stacey} H.~R.,   {Powell} D.,  2021, \mn@doi [\mnras] {10.1093/mnras/stab2295}, \href {https://ui.adsabs.harvard.edu/abs/2021MNRAS.507.3952R} {507, 3952}

\bibitem[\protect\citeauthoryear{{Robertson}}{{Robertson}}{2022}]{Robertson2022}
{Robertson} B.~E.,  2022, \mn@doi [\araa] {10.1146/annurev-astro-120221-044656}, \href {https://ui.adsabs.harvard.edu/abs/2022ARA&A..60..121R} {60, 121}

\bibitem[\protect\citeauthoryear{{Roman-Oliveira}, {Fraternali}  \& {Rizzo}}{{Roman-Oliveira} et~al.}{2023}]{Roman-Olivera+23}
{Roman-Oliveira} F.,  {Fraternali} F.,   {Rizzo} F.,  2023, \mn@doi [\mnras] {10.1093/mnras/stad530}, \href {https://ui.adsabs.harvard.edu/abs/2023MNRAS.521.1045R} {521, 1045}

\bibitem[\protect\citeauthoryear{{Romano} et~al.,}{{Romano} et~al.}{2022}]{Romano+22}
{Romano} M.,  et~al., 2022, \mn@doi [\aap] {10.1051/0004-6361/202142265}, \href {https://ui.adsabs.harvard.edu/abs/2022A&A...660A..14R} {660, A14}

\bibitem[\protect\citeauthoryear{{Schaerer}, {Meynet}, {Maeder}  \& {Schaller}}{{Schaerer} et~al.}{1993}]{Schaerer+93}
{Schaerer} D.,  {Meynet} G.,  {Maeder} A.,   {Schaller} G.,  1993, \aaps, \href {https://ui.adsabs.harvard.edu/abs/1993A&AS...98..523S} {98, 523}

\bibitem[\protect\citeauthoryear{{Schaerer} et~al.,}{{Schaerer} et~al.}{2020}]{Schaerer}
{Schaerer} D.,  et~al., 2020, \mn@doi [\aap] {10.1051/0004-6361/202037617}, \href {https://ui.adsabs.harvard.edu/abs/2020A&A...643A...3S} {643, A3}

\bibitem[\protect\citeauthoryear{{Schaerer}, {Marques-Chaves}, {Barrufet}, {Oesch}, {Izotov}, {Naidu}, {Guseva}  \& {Brammer}}{{Schaerer} et~al.}{2022}]{Schaerer+22}
{Schaerer} D.,  {Marques-Chaves} R.,  {Barrufet} L.,  {Oesch} P.,  {Izotov} Y.~I.,  {Naidu} R.,  {Guseva} N.~G.,   {Brammer} G.,  2022, \mn@doi [\aap] {10.1051/0004-6361/202244556}, \href {https://ui.adsabs.harvard.edu/abs/2022A&A...665L...4S} {665, L4}

\bibitem[\protect\citeauthoryear{{Scoville} et~al.,}{{Scoville} et~al.}{2007}]{Scoville+07}
{Scoville} N.,  et~al., 2007, \mn@doi [\apjs] {10.1086/516585}, \href {https://ui.adsabs.harvard.edu/abs/2007ApJS..172....1S} {172, 1}

\bibitem[\protect\citeauthoryear{{Sommovigo}, {Ferrara}, {Carniani}, {Zanella}, {Pallottini}, {Gallerani}  \& {Vallini}}{{Sommovigo} et~al.}{2021}]{Sommovigo+21}
{Sommovigo} L.,  {Ferrara} A.,  {Carniani} S.,  {Zanella} A.,  {Pallottini} A.,  {Gallerani} S.,   {Vallini} L.,  2021, \mn@doi [\mnras] {10.1093/mnras/stab720}, \href {https://ui.adsabs.harvard.edu/abs/2021MNRAS.503.4878S} {503, 4878}

\bibitem[\protect\citeauthoryear{{Stacey}, {Hailey-Dunsheath}, {Ferkinhoff}, {Nikola}, {Parshley}, {Benford}, {Staguhn}  \& {Fiolet}}{{Stacey} et~al.}{2010}]{stacey2010}
{Stacey} G.~J.,  {Hailey-Dunsheath} S.,  {Ferkinhoff} C.,  {Nikola} T.,  {Parshley} S.~C.,  {Benford} D.~J.,  {Staguhn} J.~G.,   {Fiolet} N.,  2010, \mn@doi [\apj] {10.1088/0004-637X/724/2/957}, \href {https://ui.adsabs.harvard.edu/abs/2010ApJ...724..957S} {724, 957}

\bibitem[\protect\citeauthoryear{{Stanway} \& {Eldridge}}{{Stanway} \& {Eldridge}}{2018}]{stanway}
{Stanway} E.~R.,  {Eldridge} J.~J.,  2018, \mn@doi [\mnras] {10.1093/mnras/sty1353}, \href {https://ui.adsabs.harvard.edu/abs/2018MNRAS.479...75S} {479, 75}

\bibitem[\protect\citeauthoryear{{Stanway}, {Eldridge}, {Greis}, {Davies}, {Wilkins}  \& {Bremer}}{{Stanway} et~al.}{2014}]{Stanaway+14}
{Stanway} E.~R.,  {Eldridge} J.~J.,  {Greis} S. M.~L.,  {Davies} L. J.~M.,  {Wilkins} S.~M.,   {Bremer} M.~N.,  2014, \mn@doi [\mnras] {10.1093/mnras/stu1682}, \href {https://ui.adsabs.harvard.edu/abs/2014MNRAS.444.3466S} {444, 3466}

\bibitem[\protect\citeauthoryear{{Steidel}}{{Steidel}}{2014}]{Steidel+14}
{Steidel} C.,  2014, {KBSS-MOSFIRE: Deep Near-IR Spectroscopy of High-z Galaxies in the Keck Baryonic Structure Survey Fields}, Keck Observatory Archive MOSFIRE, id.C223M

\bibitem[\protect\citeauthoryear{Tacchella et~al.,}{Tacchella et~al.}{2023}]{Tacchella+23}
Tacchella S.,  et~al., 2023, \mn@doi [Monthly Notices of the Royal Astronomical Society] {10.1093/mnras/stad1408}, 522, 6236–6249

\bibitem[\protect\citeauthoryear{{Tacconi}, {Genzel}  \& {Sternberg}}{{Tacconi} et~al.}{2020}]{Tacconi+20}
{Tacconi} L.~J.,  {Genzel} R.,   {Sternberg} A.,  2020, \mn@doi [\araa] {10.1146/annurev-astro-082812-141034}, \href {https://ui.adsabs.harvard.edu/abs/2020ARA&A..58..157T} {58, 157}

\bibitem[\protect\citeauthoryear{{{\"U}bler} et~al.,}{{{\"U}bler} et~al.}{2023}]{Ubler+23}
{{\"U}bler} H.,  et~al., 2023, \mn@doi [\aap] {10.1051/0004-6361/202346137}, \href {https://ui.adsabs.harvard.edu/abs/2023A&A...677A.145U} {677, A145}

\bibitem[\protect\citeauthoryear{{Vallini}, {Gallerani}, {Ferrara}, {Pallottini}  \& {Yue}}{{Vallini} et~al.}{2015}]{Vallini+15}
{Vallini} L.,  {Gallerani} S.,  {Ferrara} A.,  {Pallottini} A.,   {Yue} B.,  2015, \mn@doi [\apj] {10.1088/0004-637X/813/1/36}, \href {https://ui.adsabs.harvard.edu/abs/2015ApJ...813...36V} {813, 36}

\bibitem[\protect\citeauthoryear{{Vallini}, {Ferrara}, {Pallottini}, {Carniani}  \& {Gallerani}}{{Vallini} et~al.}{2021}]{Vallini+21}
{Vallini} L.,  {Ferrara} A.,  {Pallottini} A.,  {Carniani} S.,   {Gallerani} S.,  2021, \mn@doi [\mnras] {10.1093/mnras/stab1674}, \href {https://ui.adsabs.harvard.edu/abs/2021MNRAS.505.5543V} {505, 5543}

\bibitem[\protect\citeauthoryear{{Vallini} et~al.,}{{Vallini} et~al.}{2024}]{Vallini24}
{Vallini} L.,  et~al., 2024, \mn@doi [\mnras] {10.1093/mnras/stad3150}, \href {https://ui.adsabs.harvard.edu/abs/2024MNRAS.527...10V} {527, 10}

\bibitem[\protect\citeauthoryear{{Vanderhoof} et~al.,}{{Vanderhoof} et~al.}{2022}]{Vanderhoof+22}
{Vanderhoof} B.~N.,  et~al., 2022, \mn@doi [\mnras] {10.1093/mnras/stac071}, \href {https://ui.adsabs.harvard.edu/abs/2022MNRAS.511.1303V} {511, 1303}

\bibitem[\protect\citeauthoryear{{Venturi} et~al.,}{{Venturi} et~al.}{2024}]{venturi2024}
{Venturi} G.,  et~al., 2024, \mn@doi [\aap] {10.1051/0004-6361/202449855}, \href {https://ui.adsabs.harvard.edu/abs/2024A&A...691A..19V} {691, A19}

\bibitem[\protect\citeauthoryear{{Weingartner} \& {Draine}}{{Weingartner} \& {Draine}}{2001}]{WeingartnerDraine+01}
{Weingartner} J.~C.,  {Draine} B.~T.,  2001, \mn@doi [\apj] {10.1086/324035}, \href {https://ui.adsabs.harvard.edu/abs/2001ApJ...563..842W} {563, 842}

\bibitem[\protect\citeauthoryear{{Weingartner}, {Draine}  \& {Barr}}{{Weingartner} et~al.}{2006}]{Weingartner+06}
{Weingartner} J.~C.,  {Draine} B.~T.,   {Barr} D.~K.,  2006, \mn@doi [\apj] {10.1086/504420}, \href {https://ui.adsabs.harvard.edu/abs/2006ApJ...645.1188W} {645, 1188}

\bibitem[\protect\citeauthoryear{{Wolfire}, {Vallini}  \& {Chevance}}{{Wolfire} et~al.}{2022}]{Wolfire22}
{Wolfire} M.~G.,  {Vallini} L.,   {Chevance} M.,  2022, \mn@doi [\araa] {10.1146/annurev-astro-052920-010254}, \href {https://ui.adsabs.harvard.edu/abs/2022ARA&A..60..247W} {60, 247}

\bibitem[\protect\citeauthoryear{{Xiao}, {Stanway}  \& {Eldridge}}{{Xiao} et~al.}{2018}]{Xiao+18}
{Xiao} L.,  {Stanway} E.~R.,   {Eldridge} J.~J.,  2018, \mn@doi [\mnras] {10.1093/mnras/sty646}, \href {https://ui.adsabs.harvard.edu/abs/2018MNRAS.477..904X} {477, 904}

\bibitem[\protect\citeauthoryear{{Xiao}, {Galbany}, {Eldridge}  \& {Stanway}}{{Xiao} et~al.}{2019}]{Xiao+19}
{Xiao} L.,  {Galbany} L.,  {Eldridge} J.~J.,   {Stanway} E.~R.,  2019, \mn@doi [\mnras] {10.1093/mnras/sty2557}, \href {https://ui.adsabs.harvard.edu/abs/2019MNRAS.482..384X} {482, 384}

\bibitem[\protect\citeauthoryear{{Yang} et~al.,}{{Yang} et~al.}{2023}]{Yang+23}
{Yang} G.,  et~al., 2023, \mn@doi [\apjl] {10.3847/2041-8213/acd639}, \href {https://ui.adsabs.harvard.edu/abs/2023ApJ...950L...5Y} {950, L5}

\bibitem[\protect\citeauthoryear{{Zanella} et~al.,}{{Zanella} et~al.}{2018}]{Zanella}
{Zanella} A.,  et~al., 2018, \mn@doi [\mnras] {10.1093/mnras/sty2394}, \href {https://ui.adsabs.harvard.edu/abs/2018MNRAS.481.1976Z} {481, 1976}

\bibitem[\protect\citeauthoryear{{van Hoof}, {Weingartner}, {Martin}, {Volk}  \& {Ferland}}{{van Hoof} et~al.}{2004}]{vanHoof+04}
{van Hoof} P.~A.~M.,  {Weingartner} J.~C.,  {Martin} P.~G.,  {Volk} K.,   {Ferland} G.~J.,  2004, \mn@doi [\mnras] {10.1111/j.1365-2966.2004.07734.x}, \href {https://ui.adsabs.harvard.edu/abs/2004MNRAS.350.1330V} {350, 1330}

\makeatother
\end{thebibliography}

\onecolumn
\begin{appendix}
\section{Exponential declining SF}\label{sec:exp_decl}
\begin{figure}[htbp]
    \centering
    \resizebox{0.65\hsize}{!}{\includegraphics{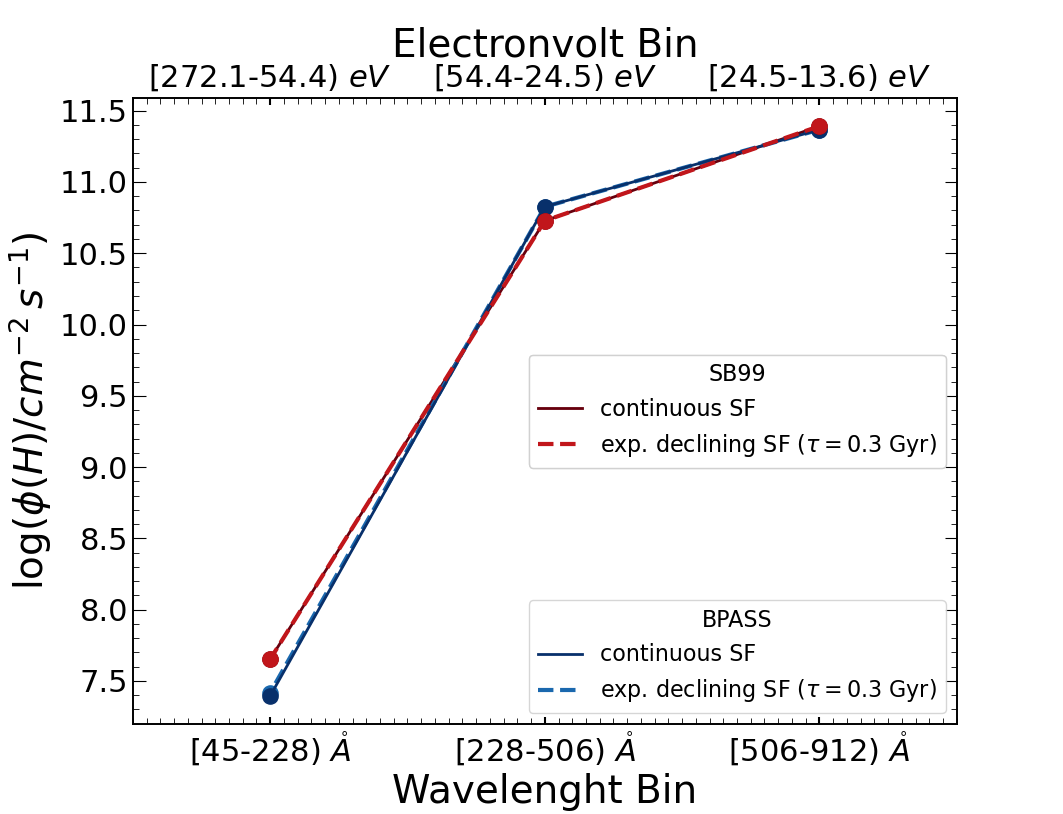}}
    \caption{The incident ionising flux $\phi(H)$ per energy bin/wavelength bin (\textsc{CLOUDY} models $U=-1.00$, $n=2$, $Z=0.55\, Z_{\odot}$) for BPASS (blue lines) and SB99 (red lines). The different SFH models are represented with solid (continuous SF) and dashed (exponential declining SF $\tau=0.3\, \rm Gyr$) lines, respectively.}
    \label{fig:phi}
\end{figure}

\section{CIII], [OIII] and H\texorpdfstring{$\alpha$}{a} SB99 predictions}\label{sec:sb99predict3l}

\begin{table*}[htbp]
\centering
\caption{The slope $\alpha$ and the intercept $\beta$ of the linear regression for the CIII] $1909\mathring{A}$, [OIII] $5007\rm \text{\AA}$ and H$\alpha$ lines.}     
\label{table:RESULTS_2}
\resizebox{0.9\textwidth}{!}{
\begin{tabular}{cccccccc}
\toprule
\toprule
\multirow{2}{*}{SFH} & \multirow{2}{*}{$k_s$} & \multicolumn{2}{c}{CIII{]} $1909\rm \text{\AA}$} & \multicolumn{2}{c}{{[}OIII{]} $5007\rm \text{\AA}$} & \multicolumn{2}{c}{H$\alpha$}       \\
                               &       & $\alpha$                              & $\beta$                              & $\alpha$                               & $\beta$                                & $\alpha$                     & $\beta$                      \\
\midrule
\multirow{4}{*}{continous SF}    & 0.34 "fiducial"     & $0.53_{-0.21}^{0.22}$ & $39.70_{-0.38}^{0.38}$ & $0.38_{-0.21}^{0.21}$ & $40.72_{-0.38}^{0.38}$ & $0.84_{-0.21}^{0.21}$ & $41.33_{-0.37}^{0.38}$ \\[0.08cm]
                                 & 1                   & $0.66_{-0.21}^{0.21}$ & $39.76_{-0.38}^{0.38}$ & $0.52_{-0.21}^{0.21}$ & $41.70_{-0.38}^{0.37}$ & $0.84_{-0.22}^{0.21}$ & $41.56_{-0.38}^{0.37}$ \\[0.08cm]
                                 & 5                   & $0.66_{-0.21}^{0.21}$ & $40.56_{-0.37}^{0.38}$ & $0.87_{-0.21}^{0.21}$ & $41.93_{-0.38}^{0.37}$ & $0.90_{-0.22}^{0.21}$ & $41.70_{-0.37}^{0.38}$ \\[0.08cm]
                                 & 10                  & $0.87_{-0.21}^{0.21}$ & $40.21_{-0.37}^{0.37}$ & $0.88_{-0.21}^{0.21}$ & $42.12_{-0.37}^{0.37}$ & $0.95_{-0.21}^{0.21}$ & $41.84_{-0.37}^{0.38}$ \\[0.08cm]
\midrule
\multirow{4}{*}{quenching 5 Myr} & 0.34 "fiducial"     & $0.34_{-0.21}^{0.22}$ & $38.17_{-0.38}^{0.38}$ & $0.56_{-0.21}^{0.21}$ & $37.82_{-0.37}^{0.38}$ & $0.85_{-0.21}^{0.21}$ & $41.27_{-0.38}^{0.38}$ \\[0.08cm]
                                 & 1                   & $0.47_{-0.21}^{0.21}$ & $39.35_{-0.37}^{0.37}$ & $0.42_{-0.22}^{0.22}$ & $39.34_{-0.38}^{0.39}$ & $0.84_{-0.21}^{0.21}$ & $41.53_{-0.38}^{0.37}$ \\[0.08cm]
                                 & 5                   & $0.58_{-0.21}^{0.21}$ & $39.26_{-0.37}^{0.37}$ & $0.76_{-0.21}^{0.21}$ & $40.22_{-0.38}^{0.38}$ & $1.08_{-0.21}^{0.21}$ & $41.20_{-0.38}^{0.38}$ \\[0.08cm]
                                 & 10                  & $0.85_{-0.22}^{0.22}$ & $38.76_{-0.38}^{0.38}$ & $0.69_{-0.21}^{0.21}$ & $40.71_{-0.37}^{0.38}$ & $0.97_{-0.21}^{0.21}$ & $41.34_{-0.38}^{0.38}$\\[0.08cm]
\bottomrule
\end{tabular}}
\parbox{\textwidth}{\tablefoot{Linear regression $\log(L_{line}/\rm erg\, s^{-1})=\alpha\log(\rm SFR/M_{\odot}\, yr^{-1})+\beta$ computed with a Monte Carlo fitting technique.
The $\alpha$ and $\beta$ were obtained assuming the SB99 incident spectrum.}}
\end{table*}
\newpage

\section{[NII] and \texorpdfstring{$H\beta$}{b} predictions}\label{sec:appendix}

\begin{figure*}[htbp]
    \centering
    \includegraphics[width=8cm]{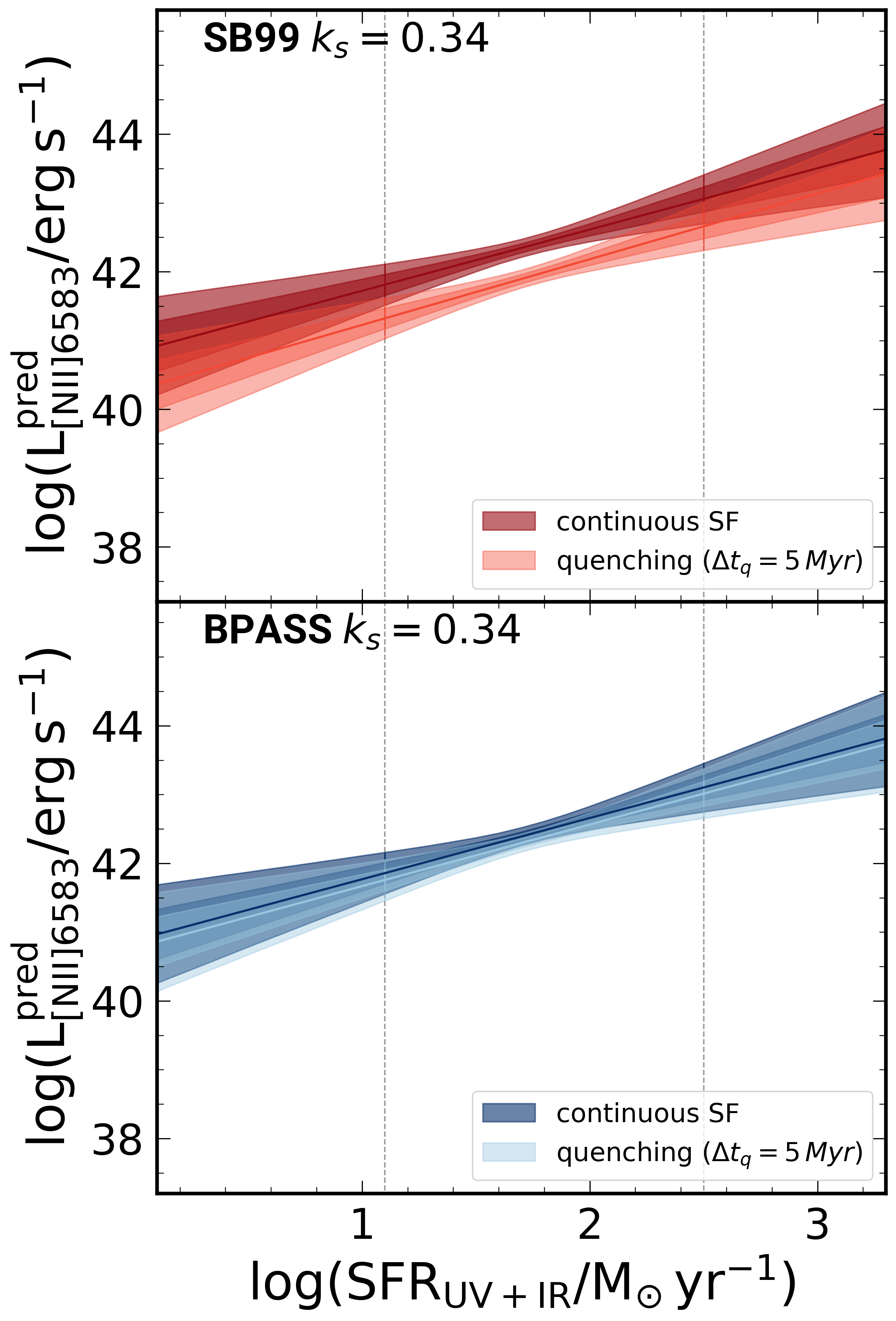}
    \includegraphics[width=8cm]{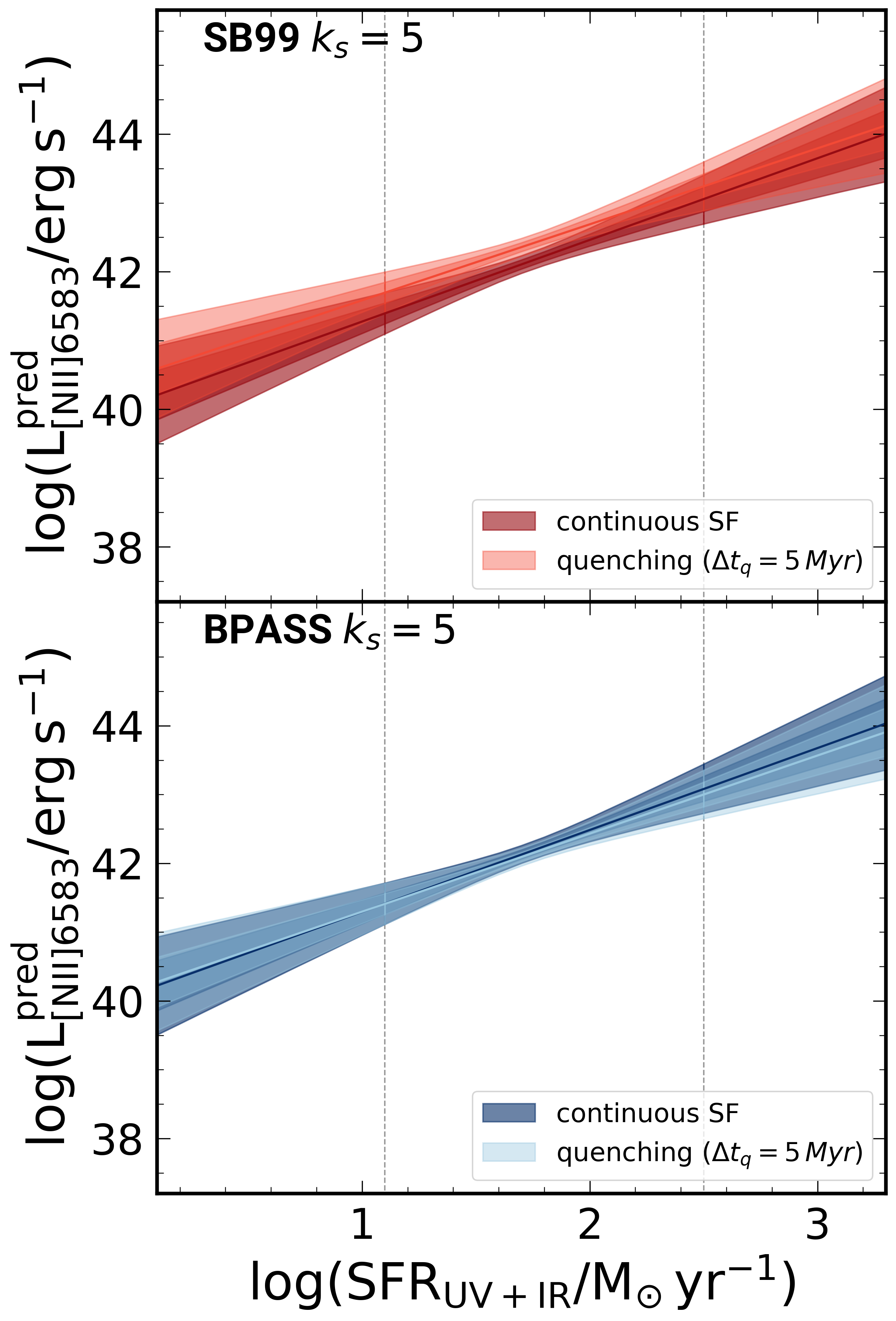}
    \caption{Predicted luminosity of [NII] $6583\rm \text{\AA}$ as a function of the total SFR (SFR$_{\rm UV+IR}$), for $k_s=0.34$ (left panel) and $k_s=5$ (right panel). The colours correspond to 1-2$\sigma$ dispersion. The central region between the grey dashed lines represents the SFR range of the ALPINE subsample modelled in this work. The red and blue colour palettes represent the two star formation histories considered in this work: continuous and $\Delta t_q=5\, \rm Myr$.}
    \label{fig:[NII]}
\end{figure*}

\begin{figure*}[htbp]
    \centering
    \includegraphics[width=8cm]{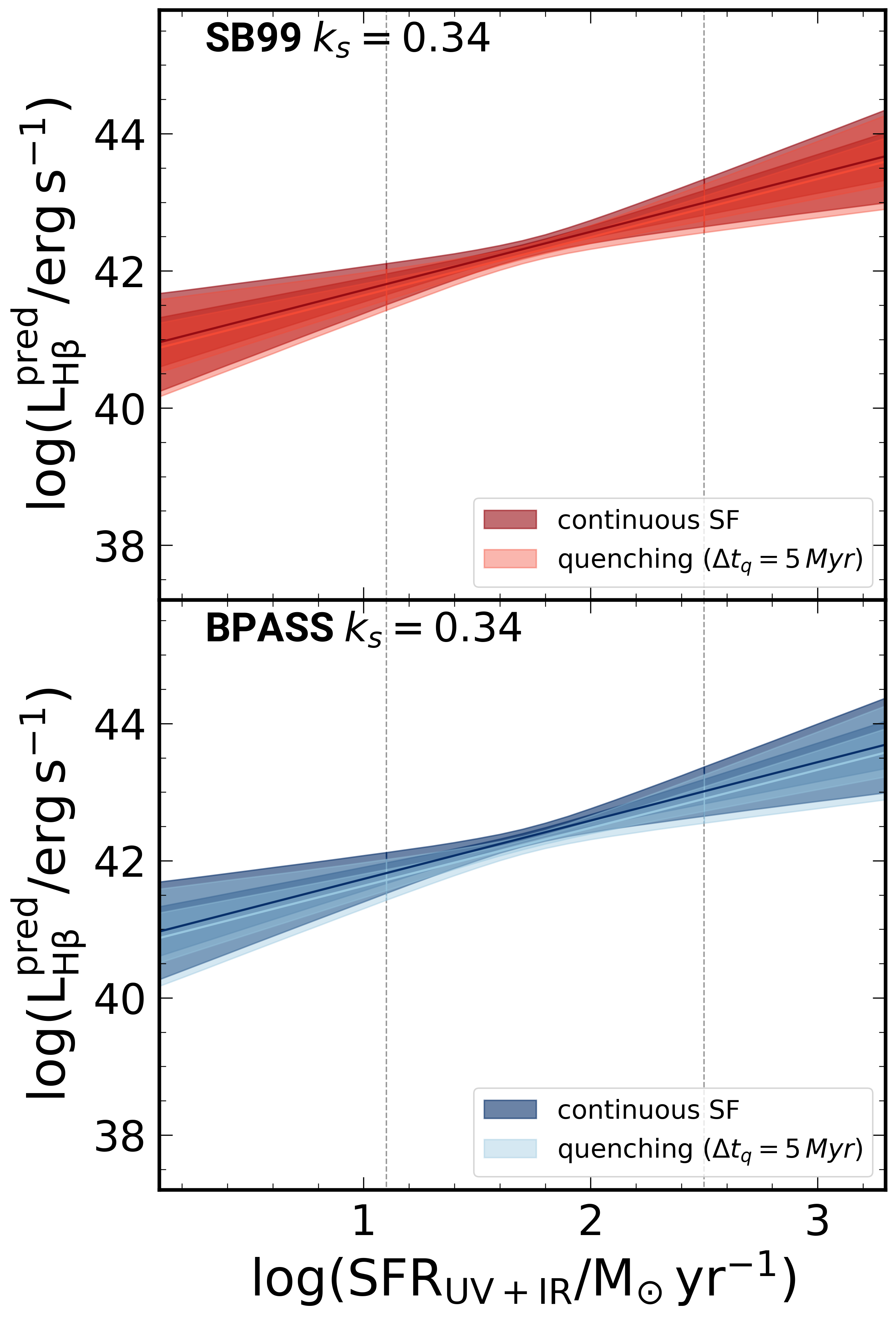}
    \includegraphics[width=8cm]{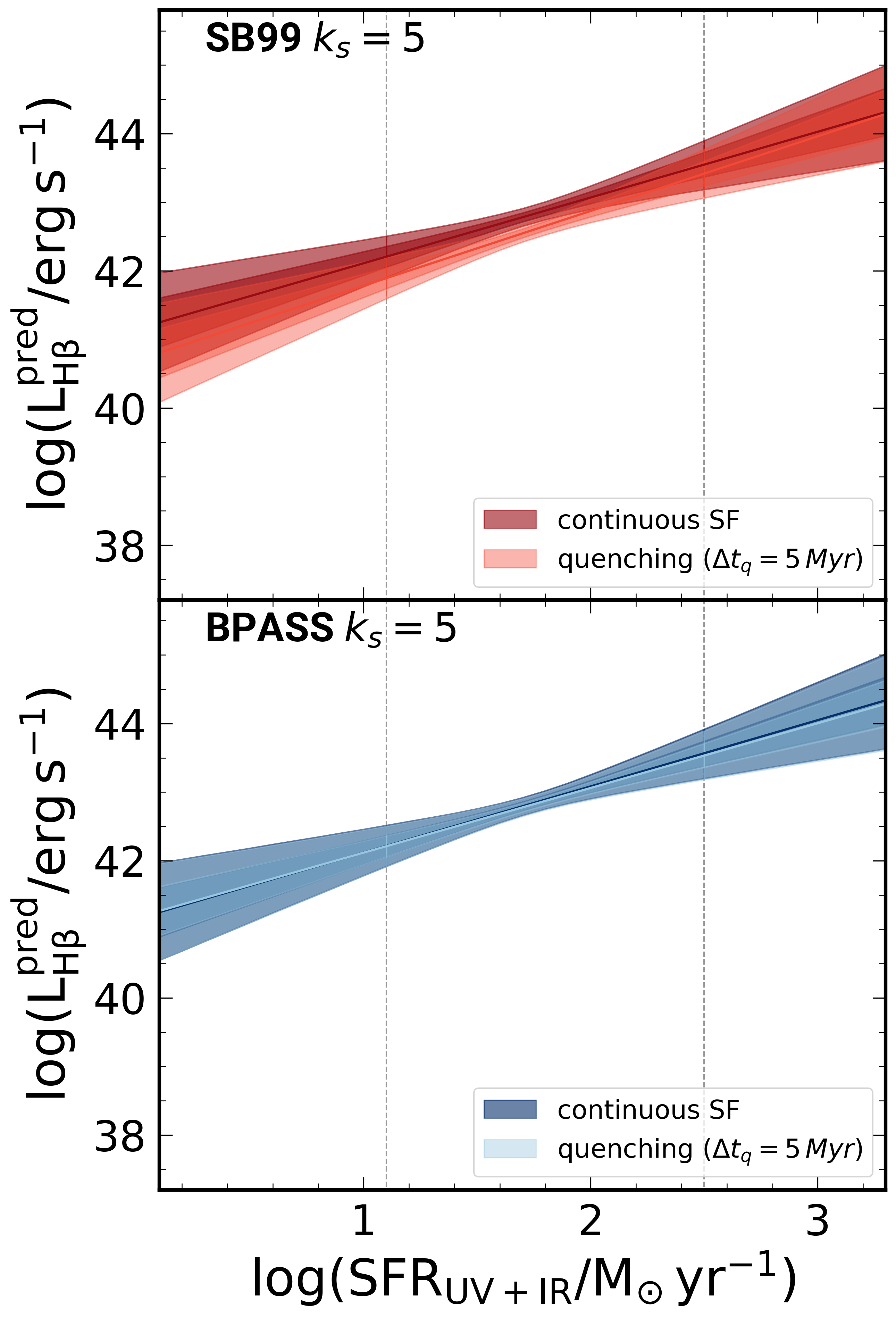}
    \caption{Predicted luminosity of $H\beta$ as a function of the total SFR (SFR$_{\rm UV+IR}$), for $k_s=0.34$ (left panel) and $k_s=5$ (right panel). The colours correspond to 1-2$\sigma$ dispersion. The central region between the grey dashed lines represents the SFR range of the ALPINE subsample modelled in this work. The red and blue colour palettes represent the two star formation histories considered in this work: continuous and $\Delta t_q=5\, \rm Myr$.}
    \label{fig:Hb}
\end{figure*}

\begin{table*}[htbp]
\centering
\caption{The slope $\alpha$ and the intercept $\beta$ of the linear regression for the [NII]$6583\rm \text{\AA}$ and H$\beta$ lines.}
\resizebox{\textwidth}{!}{
\begin{tabular}{cccccccccc}
\toprule
\toprule
\multirow{3}{*}{SFH} & \multirow{3}{*}{$k_s$} & \multicolumn{4}{c}{$H\beta$}                                                                      & \multicolumn{4}{c}{{[}NII{]} $6583\rm \text{\AA}$}                                          \\
                                                &                        & \multicolumn{2}{c}{SB99}                        & \multicolumn{2}{c}{BPASS}                       & \multicolumn{2}{c}{SB99}                       & \multicolumn{2}{c}{BPASS}                      \\
                                                &                        & $\alpha$                     & $\beta$                       & $\alpha$                      & $\beta$                      & $\alpha$                     & $\beta$                      & $\alpha$                     & $\beta$                      \\
\midrule
\multirow{4}{*}{continous SF}    & 0.34 "f."     & $0.84_{-0.21}^{0.21}$ & $40.87_{-0.38}^{0.38}$ & $0.85_{-0.21}^{0.21}$ & $40.88_{-0.37}^{0.38}$ & $0.89_{-0.22}^{0.22}$ & $40.83_{-0.39}^{0.38}$ & $0.88_{-0.21}^{0.21}$ & $40.87_{-0.38}^{0.38}$ \\[0.08cm]
                                 & 1                   & $0.85_{-0.22}^{0.21}$ & $41.08_{-0.38}^{0.39}$ & $0.84_{-0.22}^{0.21}$ & $41.09_{-0.38}^{0.39}$ & $0.99_{-0.21}^{0.21}$ & $40.75_{-0.38}^{0.38}$ & $0.98_{-0.22}^{0.22}$ & $40.80_{-0.38}^{0.39}$ \\[0.08cm]
                                 & 5                   & $0.95_{-0.21}^{0.21}$ & $41.15_{-0.38}^{0.37}$ & $0.96_{-0.22}^{0.21}$ & $41.15_{-0.37}^{0.38}$ & $1.18_{-0.21}^{0.21}$ & $40.08_{-0.37}^{0.38}$ & $1.19_{-0.21}^{0.22}$ & $40.10_{-0.38}^{0.38}$ \\[0.08cm]
                                 & 10                  & $0.91_{-0.21}^{0.21}$ & $41.30_{-0.38}^{0.37}$ & $0.93_{-0.21}^{0.21}$ & $41.29_{-0.37}^{0.38}$ & $1.11_{-0.21}^{0.21}$ & $40.00_{-0.38}^{0.37}$ & $1.12_{-0.21}^{0.21}$ & $40.00_{-0.38}^{0.38}$ \\[0.08cm]
\midrule
\multirow{4}{*}{quenching 5 Myr} & 0.34 "f."     & $0.84_{-0.21}^{0.21}$ & $40.79_{-0.38}^{0.38}$ & $0.84_{-0.22}^{0.21}$ & $40.78_{-0.38}^{0.38}$ & $0.95_{-0.21}^{0.21}$ & $40.27_{-0.38}^{0.37}$ & $0.89_{-0.21}^{0.21}$ & $40.77_{-0.38}^{0.38}$ \\[0.08cm]
                                 & 1                   & $0.85_{-0.22}^{0.21}$ & $41.03_{-0.37}^{0.39}$ & $0.81_{-0.21}^{0.21}$ & $41.08_{-0.38}^{0.38}$ & $0.89_{-0.21}^{0.21}$ & $40.72_{-0.38}^{0.38}$ & $0.96_{-0.21}^{0.21}$ & $40.75_{-0.38}^{0.38}$ \\[0.08cm]
                                 & 5                   & $1.08_{-0.21}^{0.22}$ & $40.69_{-0.38}^{0.37}$ & $0.94_{-0.21}^{0.21}$ & $41.17_{-0.38}^{0.37}$ & $1.10_{-0.21}^{0.21}$ & $40.48_{-0.38}^{0.38}$ & $1.13_{-0.21}^{0.21}$ & $40.16_{-0.38}^{0.38}$ \\[0.08cm]
                                 & 10                  & $0.97_{-0.21}^{0.21}$ & $40.83_{-0.39}^{0.37}$ & $0.91_{-0.21}^{0.21}$ & $41.32_{-0.38}^{0.38}$ & $1.01_{-0.21}^{0.22}$ & $40.55_{-0.39}^{0.38}$ & $1.09_{-0.21}^{0.21}$ & $40.09_{-0.37}^{0.37}$\\[0.08cm]
\bottomrule
\end{tabular}}
\parbox{\textwidth}{\tablefoot{Linear regression $\log(L/\rm erg\, s^{-1})=\alpha\log(\rm SFR/M_{\odot}\, yr^{-1})+\beta$ computed with a Monte Carlo fitting technique.}}
\end{table*}
\end{appendix}
\label{LastPage}
\end{document}